# pcNN-$S_N$: a physics-enhanced predictor-corrector neural network algorithm coupled with the discrete ordinates method for neutron transport equations

Cong Liu[a,b], Junxia Wei[a,b], Shuai Wang[a], Heng Yong[a,b]*

a. Institute of Applied Physics and Computational Mathematics, Beijing 100094, China

b. National Key Laboratory of Computational Physics, Beijing 100088, China

*Corresponding author.

liu_cong@iapcm.ac.cn (C. Liu),

wei_junxia@iapcm.ac.cn (J. Wei),

wang_shuai@iapcm.ac.cn (S. Wang),

yong_heng@iapcm.ac.cn (H. Yong).

# Abstract

Super-resolution machine learning methods hold promise for addressing the high computational costs associated with high-dimensional features in traditional transport simulations. This paper proposes a physics-enhanced neural network coupling algorithm for discrete ordinates ($S_N$) transport calculations. Based on high- and low-accuracy data, we develop a predictor-corrector neural network (pcNN) to predict correction factors for coarse-grid solutions. Based on transport equations, local physical features containing subgrid-scale information are selected as network inputs. Dimensionless transformation of physical quantities reduces the input space dimensionality and narrows parameter distributions, facilitating network training. Physical constraints are incorporated into the network model, and the corrected solution retains its original conservation properties. pcNN and $S_N$ transport sweeps are internally coupled, and numerical solutions are corrected online. Numerical results demonstrate that this algorithm can effectively improve coarse-grid solutions to achieve a level of accuracy comparable to that of fine-grid results. The physics-based network design enables the model to demonstrate a certain degree of reusability under different computational conditions. Compared to mesh refinement, the pcNN-$S_N$ algorithm reduces the computation time for the three-dimensional test problem by more than 60%.

## 1. Introduction

The motion and collisions of microscopic particles such as neutrons, photons, and electrons in a medium can be described by the Boltzmann transport equation. Numerical simulation of transport equations constitutes a core component of neutronics calculations in nuclear engineering. To ensure simulation reliability, selecting an appropriate numerical discretization scheme and mesh size is crucial for deterministic approaches such as the discrete ordinates ($S_N$) method [1]. Engineering problems typically exhibit significant spatial heterogeneity. If the grid optical thickness is too large, the $S_N$ method may generate large spatial discretization errors. Given the high-dimensional nature of transport equations involving space, angles, and energy, the computational cost of fine-mesh simulations is often prohibitively high. Therefore, achieving efficient and reliable transport simulations at the lowest possible computational cost has long been a major research focus in the field of particle transport.

In recent years, machine learning methods based on neural networks have garnered significant attention due to their powerful approximation capabilities and adaptability to high-dimensional problems. As a novel type of partial differential equation solver, physics-informed neural networks (PINNs) employ equation residuals as the optimization objective for training networks and have been applied to solve particle diffusion and transport equations [2-4]. The classical PINNs method has difficulties when computing complex transport problems, leading to the development of algorithms such as multiscale loss functions [5], Fourier features [6,7], data augmentation [8,9], and interface treatment [10-12]. However, current PINNs still face challenges in terms of computational accuracy and efficiency when addressing practical neutronics problems.

At the same time, data-driven machine learning methods are increasingly being applied to enhance traditional numerical calculations. By substituting portions of conventional parameter generation or simulation processes, they improve the modeling accuracy or computational efficiency of localized procedures. Martin et al. employed neural networks to achieve direct prediction of state parameters to material cross sections [13], which serve as the fundamental parameters for transport simulations. Chan and Dufek integrated neural networks with PCA techniques to establish predictive models for multiple material cross sections [14]. Saller et al. optimized energy group structures using simulated annealing and random forest models [15], enabling results close to those of refined energy groups. For transport calculations, Tano and Ragusa proposed a Sweep-Net surrogate model to replace Gaussian elimination computations in the $S_N$-discontinuous finite element method [16], thereby accelerating the local solution process. Huang et al. proposed a machine learning moment closure model [17,18] for radiative transport, which exhibits lower errors compared to traditional $P_N$ closure relations. Whewell and McClarren developed a neural network model directly mapping neutron flux to source terms [19], eliminating the need to store multi-group fission cross sections and scattering matrices. Most of these studies focus on improving computational efficiency, and it is generally difficult to strictly preserve the original physical properties.

For applications requiring iterative numerical simulations of specific computational models, data-driven methods can serve as a comprehensive alternative to traditional numerical methods, supporting related design optimization and uncertainty analysis. This type of research uses massive simulation data and machine learning methods to establish a mapping relationship between macroscopic physical parameters and target quantities or distribution fields [20]. Once sufficient training data has been collected and a surrogate model has been developed, traditional numerical simulations will be completely replaced. Shriver et al. developed a neural network architecture to predict the normalized power and effective multiplication factor (keff) of a two-dimensional assembly model [21]. Gong et al. combined reduced-order models with k-nearest-neighbors and decision trees to predict neutron flux and power distribution [22]. Foad et al. combined reduced-order models with deep neural networks to approximate time-varying transient behavior [23]. Ren et al. developed a model for predicting the keff using a long short-term memory network [24]. Bei et al. used a backpropagation neural network to rapidly predict keff and channel-level neutron flux distributions using core operating parameters as input [25]. Similarly, some studies used other network architectures or machine learning methods in place of transport simulations to predict the neutron physical quantities of interest [26-29]. These machine learning models can overcome the computational costs associated with traditional numerical calculations and enable rapid predictions, but they often lack physical interpretability and are difficult to directly generalize to other computational models.

In recent years, the concepts behind super-resolution machine learning methods in computer vision [30] have been applied to improve traditional numerical simulations, using low-resolution or low-fidelity simulation results to predict high-resolution or high-fidelity outcomes. Hanna et al. proposed a machine learning-based coarse-grid CFD approach [31]. By training network models using simulation data from both fine and coarse grids, they can predict the error distribution in coarse-grid calculations, thereby obtaining flow field distributions that closely approximate those of fine-grid calculations. In the field of neutronics, Grechanuk et al. employed machine learning methods to predict deviations between simulated and experimental values of the fission system eigenvalue [32]. Oktaviana et al. developed a machine learning method using high-fidelity Monte Carlo calculations and low-fidelity diffusion calculations [33], achieving error correction for $k_{eff}$ and power results. Akbari et al. conducted a similar machine learning study on VVER reactors [34]. Berry et al. proposed an upsampling network for Monte Carlo transport calculations [35], which takes low-resolution flux tally results as input to predict results with higher spatial and energy resolution. Based on one-dimensional calculation results, Lei et al. developed a neural network correction model for key-point dose rates [36], yielding results that closely match those from direct three-dimensional simulations. All of these studies employ post-processing methods in which low-resolution/low-fidelity simulations and neural network corrections are completely separate from one another. In this approach, neural networks typically have a one-to-one correspondence with computational models, which limits their generalization ability. Second, it is generally difficult to preserve physical properties during post-processing corrections.

To address the high cost of traditional fine-grid simulations, this paper proposes a physics-enhanced neural network coupling algorithm for $S_N$ neutron transport calculations. We employ the traditional $S_N$ difference method to generate high- and low-accuracy simulation data. Based on physical information and the constraints imposed by transport equations, we develop a predictor-corrector neural network model to correct errors in low-accuracy solutions on coarse grids. The neural network model is closely coupled with the traditional $S_N$ sweep calculation process. Using the physical field features obtained from $S_N$ sweeps as input, the neural network model performs online correction of the local numerical solution. Tests on typical transport problems show that the coupling algorithm effectively corrects coarse-grid solutions, yielding results closer to high-accuracy fine-grid results. Due to the introduction of physical feature information, the network model demonstrates excellent reusability across different cross section and grid conditions, achieving over 60% reduction in computational time for three-dimensional cases.

The innovations of this paper are as follows. Compared to related machine learning research in the transport field, our network fully incorporates physical characteristics, and the network-corrected solution strictly preserves the original particle conservation properties. We find that supplementing key physical information at the subgrid scale is essential for establishing a mapping relationship between coarse- and fine-grid solutions. To address this, we incorporate information on the local source distribution. Second, the high-dimensional parameter space and wide parameter distribution range resulting from the transport equation make training the network extremely difficult. To address this issue, we start with the transport equation and, by applying dimensionless transformations to the physical quantities, successfully reduce the dimensionality of the network's input space and narrow the range of parameter distributions. And we incorporate physical constraints into network modeling to ensure conservation of numerical solutions. These strategies enable our algorithm to achieve excellent reusability.

The remainder of this paper is organized as follows. Sec. 2 reviews the $S_N$ difference discretization of transport equations. The input parameter design of predictor-corrector neural networks is introduced in Sec. 3. Physical constraints and loss function construction are presented in Sec. 4. The pcNN-$S_N$ coupling algorithm is described in Sec. 5. Numerical verification is given in Sec 6, respectively. Finally, some conclusions are drawn in Sec. 7.

## 2. $S_N$ difference discretization

The one-group transport equation on Cartesian grids is considered in this paper. The method presented here can be directly applied to multi-group, anisotropic scattering problems. The $S_N$ transport equation in 3-D Cartesian geometry is given by:

$$\frac{1}{v}\frac{\partial \psi}{\partial t}+\mu_m\frac{\partial \psi_m}{\partial x}+\eta_m\frac{\partial \psi_m}{\partial y}+\xi_m\frac{\partial \psi_m}{\partial z}+\Sigma_t\psi_m=Q_m \tag{1}$$

where $\psi$ is the neutron angular flux, $t$ is the time variable, $x$, $y$, $z$ are the spatial variables, $\mu$, $\eta$, and $\xi$ represent the direction cosines of the angular vector relative to the coordinate axes, $v$ is the neutron velocity, $\Sigma_t$ is the total cross section, $Q$ is the total source term, and $m$ is the discrete direction index. Discretizing the time term using a backward Euler scheme and omitting the direction index yields:

$$\mu\frac{\partial \psi^n}{\partial x}+\eta\frac{\partial \psi^n}{\partial y}+\xi\frac{\partial \psi^n}{\partial z}+\left(\Sigma_t+\frac{1}{v\Delta t}\right)\psi^n=Q^n+\frac{\psi^{n-1}}{v\Delta t} \tag{2}$$

where the subscripts $n$ and $n$-1 denote the $n$-th and ($n$-1)-th time steps, respectively, $\Delta t$ is the time step size. The spatial domain is divided into Cartesian cells, and spatial discretization is performed using a finite difference scheme [37]. Integrating Eq. (2) over the grid gives:

$$\frac{|\mu|}{\Delta x}(\psi_{iout}^{n}-\psi_{iin}^{n})+\frac{|\eta|}{\Delta y}(\psi_{jout}^{n}-\psi_{jin}^{n})+\frac{|\xi|}{\Delta z}(\psi_{kout}^{n}-\psi_{kin}^{n})+\tilde{\Sigma}_{t}\bar{\psi}^{n}=\tilde{Q} \tag{3}$$

$$\tilde{\Sigma}_{t}=\left(\Sigma_{t}+\frac{1}{v\Delta t}\right) \tag{4}$$

$$\tilde{Q}=\bar{Q}^{n}+\frac{\bar{\psi}^{n-1}}{v\Delta t} \tag{5}$$

where $\Delta x$, $\Delta y$, and $\Delta z$ represent the grid step sizes along the $x$, $y$, and $z$ axes, respectively. $\psi_{iout}$ and $\psi_{iin}$ denote the average outgoing and incoming angular fluxes of the grid boundaries perpendicular to the $x$ axis, respectively, and similarly for the $y$ and $z$ axes, $\bar{\psi}$ and $\bar{Q}$ represent the average angular flux and source term of the grid cell, respectively.

To close the Eq. (3), the difference relations must be supplemented. Omitting the time step index, the weighted difference scheme's closure is

$$\bar{\psi}=\frac{1}{1+\alpha_{i}}\psi_{iout}+\frac{\alpha_{i}}{1+\alpha_{i}}\psi_{iin} \tag{6}$$

where $\alpha_i$ is the weight coefficient for the $x$ dimension, and similarly for the $y$ and $z$ dimensions. For the classic diamond scheme, $\alpha$ is set to 1. The calculation formula for $\bar{\psi}$ can be derived by substituting the closure into Eq. (3). The adaptive weighted difference scheme first employs a diamond difference assumption to provide the estimated average angular flux, then adjusts the weight coefficients to ensure numerical solutions remain non-negative. Afterwards, the angular flux of each outgoing boundary is obtained as

$$\psi_{iout}=\left(1+\alpha_{i}\right)\bar{\psi}-\alpha_{i}\psi_{iin} \tag{7}$$

This paper considers the $S_N$ equations in 1-D planar and 3-D Cartesian geometries. The discretization of 1-D equation is similar to the aforementioned derivation. For brevity, subsequent formulas are presented in the form of steady-state equation as Eq. (3). After numerical discretization, neutron flux solutions are obtained using transport sweep and source iteration algorithms.

# 3. Physical feature inputs for pcNN

## 3.1 Subgrid source distribution modeling

Existing research on super-resolution modeling and machine learning-based error correction can be viewed as utilizing neural networks to identify and construct mappings from low-resolution solutions to high-resolution solutions under various parameter conditions. In some computer vision tasks, convolutional neural networks extract multiscale features through downsampling, establishing a mapping between low-resolution features and high-resolution outputs.

To improve numerical simulations, this requires us to identify key features of low-accuracy solutions and ensure that these features have a corresponding functional relationship with the high-accuracy solutions of interest. All variables in Eq. (3) can be regarded as feature parameters of the transport solution, including the grid step size, discrete direction, material cross section, source strength, and incoming/outgoing angular flux. However, Eq. (3) contains only grid-averaged source information and does not account for variations in the source distribution within the grid. Although this does not affect the construction of the difference discretization scheme, the behavior of the fine-grid solution cannot be fully captured by these coarse-grid averaged features.

The source terms on the right-hand side of the transport equation account for external sources and scattering sources, with the scattering sources being determined by the neutron flux and the material scattering cross section. The neutron flux is typically not uniformly distributed within the grid. As shown in Figure 1, within a coarse-grid region, even when the average source strength is the same, the source distribution at the sub-grid scale may vary, resulting in different deviations in the corresponding low-accuracy solutions. Therefore, it is necessary to supplement the higher-order information of the source distribution for our machine learning model.

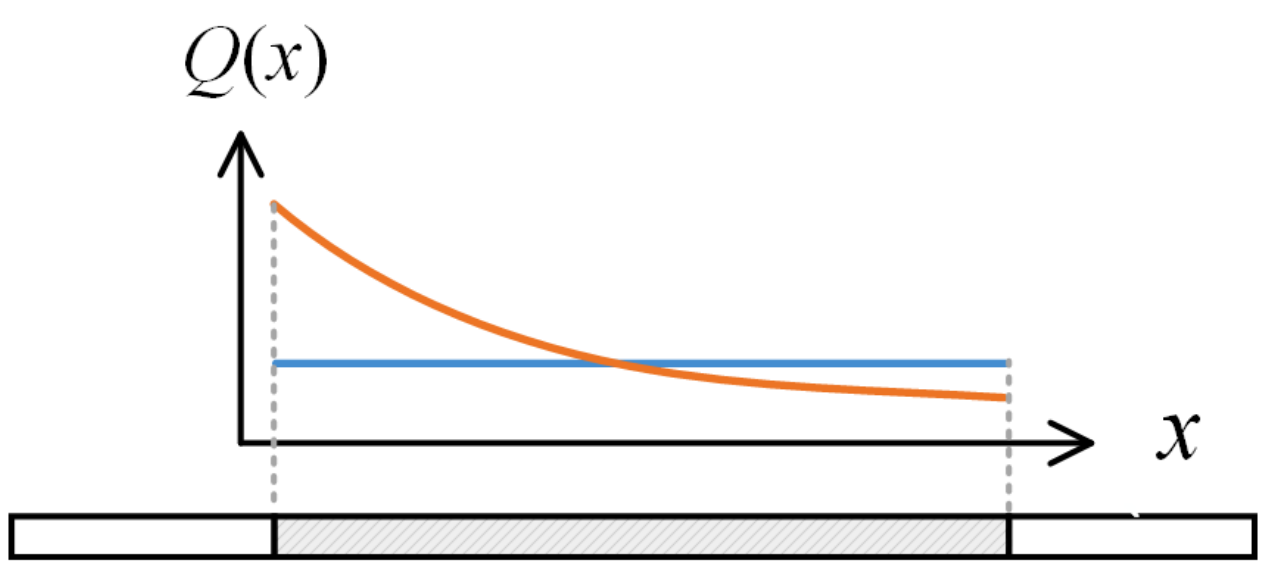

Fig. 1. Grid source strength distribution

We develop a source reconstruction algorithm compatible with the $S_N$ source iteration and sweep framework, where the reconstruction information serves as feature input for the neural network. The $S_N$ source iteration can be expressed as

$$\mathbf{L}\psi^{(l)} = \mathbf{S}\psi^{(l-1)} + q \tag{8}$$

where **L** and **S** denote the transport operator and scattering operator, respectively; $q$ represents the external source; and ($l$) indicates the iteration number. Thus, we can reconstruct higher-order source information based on the flux solution from the previous iteration. During the sweeping process, we record the boundary fluxes of the coarse grids. Currently, we consider the case of isotropic scattering, where only the boundary scalar fluxes need to be counted. Taking the x-dimension as an example,

$$\phi_{i+1/2}^{(l-1)} = \sum_{\mu_m<0} w_m \psi_{m,iin}^{(l-1)} + \sum_{\mu_m>0} w_m \psi_{m,iout}^{(l-1)} \tag{9}$$

$$\phi_{i-1/2}^{(l-1)} = \sum_{\mu_m<0} w_m \psi_{m,iout}^{(l-1)} + \sum_{\mu_m>0} w_m \psi_{m,iin}^{(l-1)} \tag{10}$$

$$\bar{\phi} = \sum_{m=1}^{M} w_m \bar{\psi} \tag{11}$$

where $\phi_{i+1/2}$ and $\phi_{i-1/2}$ represent the scalar fluxes at the positive and negative grid boundaries along the x-dimension, respectively, and $\bar{\phi}$ denotes the grid-averaged scalar flux. And similarly for the boundary scalar fluxes $\phi_{j\pm1/2}$ and $\phi_{k\pm1/2}$ in the $y$ and $z$ dimensions.

We define the interpolation basis functions as

$$N_{\alpha,\beta,\gamma}(x,y,z) = \left(1+\alpha P_1^x(x)\right)\left(1+\beta P_1^y(y)\right)\left(1+\gamma P_1^z(z)\right) \tag{12}$$

$$P_1^x(x) = 2x/\Delta x - 1 \tag{13}$$

$$P_1^y(y) = 2y/\Delta y - 1 \tag{14}$$

$$P_1^z(z) = 2z/\Delta z - 1 \tag{15}$$

where the values of $\alpha$, $\beta$, and $\gamma$ are either 1 or -1. Adopting an interpolation polynomial form, we assume the flux distribution within the grid can be expressed as

$$\phi(x,y,z) = a_0 \sum_{\alpha=\pm1}\sum_{\beta=\pm1}\sum_{\gamma=\pm1} \phi^{\alpha,\beta,\gamma} N_{\alpha,\beta,\gamma}(x,y,z) \tag{16}$$

$$\phi^{\alpha,\beta,\gamma} = \phi_{i+\alpha/2}\phi_{j+\beta/2}\phi_{k+\gamma/2} \tag{17}$$

where $a_0$ is an undetermined constant coefficient, and it can be obtained according to the following relationship:

$$\int_0^{\Delta z} dz \int_0^{\Delta y} dy \int_0^{\Delta z} \phi(x,y,z)dx = \bar{\phi} \int_0^{\Delta z} dz \int_0^{\Delta y} dy \int_0^{\Delta z} dx \tag{18}$$

Therefore, the flux distribution function is reconstructed, and its first-order moment in the x-direction can be derived as

$$\phi_x = \frac{\int_0^{\Delta z} dz \int_0^{\Delta y} dy \int_0^{\Delta x} P_1^x(x)\phi(x,y,z)dx}{\int_0^{\Delta z} dz \int_0^{\Delta y} dy \int_0^{\Delta x} dx} = \frac{\bar{\phi}\left(\phi_{i+1/2} - \phi_{i-1/2}\right)}{3\left(\phi_{i+1/2} + \phi_{i-1/2}\right)} \tag{19}$$

Accordingly, the first-order moment of the scattering source is

$$Q_x = \Sigma_s \phi_x \tag{20}$$

where $\Sigma_s$ is the scattering cross section. The first-order moments in other dimensions are constructed in the same manner.

### 3.2 Dimensionless representation of local features

The local features of coarse-grid solutions should be network inputs. Eq. (3) contains 15 variables, including 3 grid step sizes, 3 directional cosines, 1 total cross section, 1 average source strength, 3 incident angular fluxes, 3 outgoing angular fluxes, and 1 grid-averaged angular flux. In addition, the reconstructed source information includes 3 first-order moments of the source distribution. On the other hand, there are significant variations in the values of these variables, and within a single computational problem, the range of a single variable may span multiple orders of magnitude. It is unwise to use these variables directly as network inputs. The high dimensionality of the parameter space and the wide distribution range of variables pose significant challenges for neural network training. Traditional normalization methods in machine learning, such as min-max processing, cannot resolve this training issue.

Starting from the physical equation, we design the neural network inputs using dimensionless techniques. Dividing both sides of Eq. 3 by the collision reaction rate $\Sigma_t \bar{\psi}$ yields

$$\frac{|\mu|}{\Sigma_t \Delta x}\left(\frac{\psi_{iout} - \psi_{iin}}{\bar{\psi}}\right) + \frac{|\eta|}{\Sigma_t \Delta y}\left(\frac{\psi_{jout} - \psi_{jin}}{\bar{\psi}}\right) + \frac{|\xi|}{\Sigma_t \Delta z}\left(\frac{\psi_{kout} - \psi_{kin}}{\bar{\psi}}\right) + 1 = \frac{\bar{Q}}{\Sigma_t \bar{\psi}} \tag{21}$$

This transformed equation includes the optical thickness of the grid as

$$\tau_x = \frac{\Sigma_t \Delta x}{|\mu|},\ \tau_y = \frac{\Sigma_t \Delta y}{|\eta|},\ \tau_z = \frac{\Sigma_t \Delta z}{|\xi|} \tag{22}$$

They are dimensionless quantities which measure the distance a particle can travel relative to its average free path in the direction of flight. From a physics perspective, this can be understood as the average number of collisions a particle undergoes over that distance. The other variables in Eq. (21) are also dimensionless quantities, including the normalized boundary angular fluxes and the ratio of the source term to the collision term. Using the same dimensionless approach, the first-order moments of the source distribution are normalized using the grid-averaged source strength, i.e., $Q_x/\bar{Q}$, $Q_y/\bar{Q}$, and $Q_z/\bar{Q}$.

This dimensionless treatment is a normalization method that is physically interpretable. After this processing, the variables we need to consider are 3 optical thicknesses, 6 normalized boundary angular fluxes, 1 source-to-collision term ratio, and 3 normalized source first-order moments. The input parameters for pcNN are

$$\mathbf{u}_{input} = \left\{ \begin{array}{l} \dfrac{\Sigma_t \Delta x}{|\mu|}, \dfrac{\Sigma_t \Delta y}{|\eta|}, \dfrac{\Sigma_t \Delta z}{|\xi|}, \\ \dfrac{\psi_{iin}}{\bar{\psi}}, \dfrac{\psi_{jin}}{\bar{\psi}}, \dfrac{\psi_{kin}}{\bar{\psi}}, \dfrac{\psi_{iout}}{\bar{\psi}}, \dfrac{\psi_{jout}}{\bar{\psi}}, \dfrac{\psi_{kout}}{\bar{\psi}}, \\ \dfrac{\bar{Q}}{\Sigma_t \bar{\psi}}, \dfrac{Q_x}{\bar{Q}}, \dfrac{Q_y}{\bar{Q}}, \dfrac{Q_z}{\bar{Q}} \end{array} \right\} \tag{23}$$

where the number of input variables is reduced from 18 to 13, achieving dimensionality reduction in the input space. Compared to the original variables, dimensionless quantities have a narrower range of values, which also reduces the difficulty of training the network. Neutron fields within different numerical ranges may obey the same physical laws, and these dimensionless quantities correspond directly to the underlying transport mechanisms. Another advantage is that situations that are previously considered extrapolation under the original input combinations may become interpolation under the new dimensionless quantities. This greatly facilitates the reuse of network models.

## 4. Physical constraints and loss functions for pcNN

For coarse-grid domains, the pcNN model is designed to predict correction factors for the low-accuracy solutions of the coarse grid, bringing them closer to the high-accuracy simulation results on fine grids.

### 4.1 Conservation constraints

During network model construction, our objective is to ensure that the corrected numerical solution preserves the inherent conservation properties of the finite difference discretization scheme. Therefore, the loss function must account not only for data constraints but also for the physical constraints of the transport equation. From discretization equation, the following expression can be derived:

$$\begin{gathered}F(\boldsymbol{\psi}) = |\mu|(\psi_{iout} - \psi_{iin})\Delta y \Delta z + |\eta|(\psi_{jout} - \psi_{jin})\Delta x \Delta z + \\ |\xi|(\psi_{kout} - \psi_{kin}^{high})\Delta x \Delta y + \left(\Sigma_t \bar{\psi} - \bar{Q}\right)\Delta x \Delta y \Delta z\end{gathered} \tag{24}$$

This equation represents the balance between inflow, outflow, collision, and generation within the grid. Whether on coarse or fine grids, the difference solution $\boldsymbol{\psi}^{FD}$ satisfies $F(\boldsymbol{\psi}^{FD})$ equal to zero.

We now introduce neural network corrections, defining the correction factors for the grid-averaged angular flux and the outgoing angular flux respectively as

$$f_{ave} = \frac{\bar{\psi}^{high}}{\bar{\psi}^{low}}, \quad f_{iout} = \frac{\psi_{iout}^{high}}{\psi_{iout}^{low}} \tag{25}$$

and similarly for correction factors $f_{jout}$ and $f_{kout}$. The neural network prediction is

$$f_{NN} = \left\{ f_{ave}^{NN}, f_{iout}^{NN}, f_{jout}^{NN}, f_{kout}^{NN} \right\} \tag{26}$$

and the network corrected solutions are obtained as

$$\bar{\psi}^{NN} = f_{ave}^{NN} \bar{\psi}^{low}, \quad \psi_{iout}^{NN} = f_{iout}^{NN} \psi_{iout}^{low} \tag{27}$$

and similarly for boundary angular fluxes in the $y$ and $z$ dimensions. Substituting the corrected solution into Eq. (24), the prediction errors leads to

$$F(\boldsymbol{\psi}^{NN}) \neq 0 \tag{28}$$

which means that the corrected values do not satisfy the particle balance equation, i.e., they fail to meet the conservation requirement. In the transport field, most discretization schemes can guarantee the conservation property. For the machine learning method in this paper, the situation described by Eq. (28) is undesirable.

We address this conservation issue by introducing hard and soft constraints into the network model. First, the neural network is modified to predict only the outgoing angle flux correction factors as

$$f_{NN} = \left\{ f_{iout}^{NN}, f_{jout}^{NN}, f_{kout}^{NN} \right\} \tag{29}$$

The corresponding loss function is

$$loss_{out} = \frac{1}{N}\sum_{n=1}^{N}\left( (\frac{f_{iout,n}^{NN} - f_{iout,n}^{Data}}{f_{iout,n}^{Data}})^2 + (\frac{f_{jout,n}^{NN} - f_{jout,n}^{Data}}{f_{jout,n}^{Data}})^2 + (\frac{f_{kout,n}^{NN} - f_{kout,n}^{Data}}{f_{kout,n}^{Data}})^2 \right) \tag{30}$$

where $N$ denotes the total number of training samples, $f^{Data}$ and $f^{NN}$ denote the correction factors of training data and network prediction, respectively. We substitute the corrected outgoing angular fluxes into the balance equation, and the corrected solution for the grid-averaged value is directly obtained using a hard constraint approach as

$$\bar{\psi}^{fix} = \frac{\bar{Q}}{\Sigma_t} - \frac{|\mu|}{\Sigma_t \Delta x}(\psi_{iout}^{NN} - \psi_{iin}) - \frac{|\eta|}{\Sigma_t \Delta y}(\psi_{jout}^{NN} - \psi_{jin}) - \frac{|\xi|}{\Sigma_t \Delta z}(\psi_{kout}^{NN} - \psi_{kin}) \tag{31}$$

Under this treatment, the corrected solutions satisfy the conservation requirement.

However, due to the prediction error of the correction factors, the solution given by Eq. (31) does not strictly coincide with the fine-grid data. To address this, we incorporate soft constraints into the loss function based on the balance relationship. An equivalent correction factor for the grid-averaged solution is calculated as

$$f_{ave}^{fix} = \frac{\bar{\psi}^{fix}}{\bar{\psi}^{low}} = \frac{\bar{Q}}{\Sigma_t \bar{\psi}^{low}} - \frac{|\mu|(\psi_{iout}^{NN} - \psi_{iin})}{\Sigma_t \bar{\psi}^{low} \Delta x} - \frac{|\eta|(\psi_{jout}^{NN} - \psi_{jin})}{\Sigma_t \bar{\psi}^{low} \Delta y} - \frac{|\xi|(\psi_{kout}^{NN} - \psi_{kin})}{\Sigma_t \bar{\psi}^{low} \Delta z} \tag{32}$$

If $\bar{\psi}^{fix}$ approaches the fine-grid reference value, then $f_{ave}^{fix}$ similarly approaches the training data $f_{ave,n}^{Data}$ given by Eq. (25). Therefore, we add the following term to the loss function:

$$loss_{ave} = \frac{1}{N}\sum_{n=1}^{N}\left( \frac{f_{ave,n}^{fix} - f_{ave,n}^{Data}}{f_{ave,n}^{Data}} \right)^2 \tag{33}$$

### 4.2 Permutation-equivariant constraints

For orthogonal grids in Cartesian geometry, the outgoing boundary angular fluxes given by the difference scheme exhibit permutation-equivariant property with respect to spatial dimension. This property can be explained by representing the discretization scheme in the form of a response matrix:

$$\mathbf{A}_{4\times 4} \times \begin{pmatrix} \psi_{iin} \\ \psi_{jin} \\ \psi_{kin} \\ \bar{Q} \end{pmatrix} = \begin{pmatrix} \psi_{iout} \\ \psi_{jout} \\ \psi_{kout} \\ \bar{\psi} \end{pmatrix} \tag{34}$$

where $\mathbf{A}$ is the coefficient matrix determined by the difference method. When applying row transformations to Eq. (34), if the positions of input variables of different dimensions on the left-hand side are swapped, the positions of the outgoing flux solutions in each row on the right-hand side will also be swapped accordingly.

As shown in Fig. 2, let's take a two-dimensional Cartesian grid as an example. The conditions of grid spacing, discrete direction, and incoming angle flux are

$$\begin{cases} \Delta h_1 = (\Delta x, \Delta y) \\ \vec{\Omega}_1 = (\mu, \eta) \\ \psi_1^{in} = (\psi_{iin}, \psi_{jin}) \end{cases} \tag{35}$$

and the outgoing solution is obtained as $\psi_1^{out} = (\psi_{iout}, \psi_{jout})$. Next, rotate the grid and discrete directions to obtain the new $x$ and $y$ dimension inputs as follows:

$$\begin{cases} \Delta h_2 = (\Delta x', \Delta y') = (\Delta y, \Delta x) \\ \vec{\Omega}_2 = (\mu', \eta') = (-\eta, \mu) \\ \psi_2^{in} = (\psi'_{iin}, \psi'_{jin}) = (\psi_{jin}, \psi_{iin}) \end{cases} . \tag{36}$$

The new outgoing solution $\psi_2^{out} = (\psi'_{iout}, \psi'_{jout})$ satisfies

$$\begin{aligned} \psi'_{iout} &= \psi_{jout} \\ \psi'_{jout} &= \psi_{iout} \end{aligned} \tag{37}$$

This is due to the symmetry of the leakage operator across all dimensions and the orthogonality of the grid.

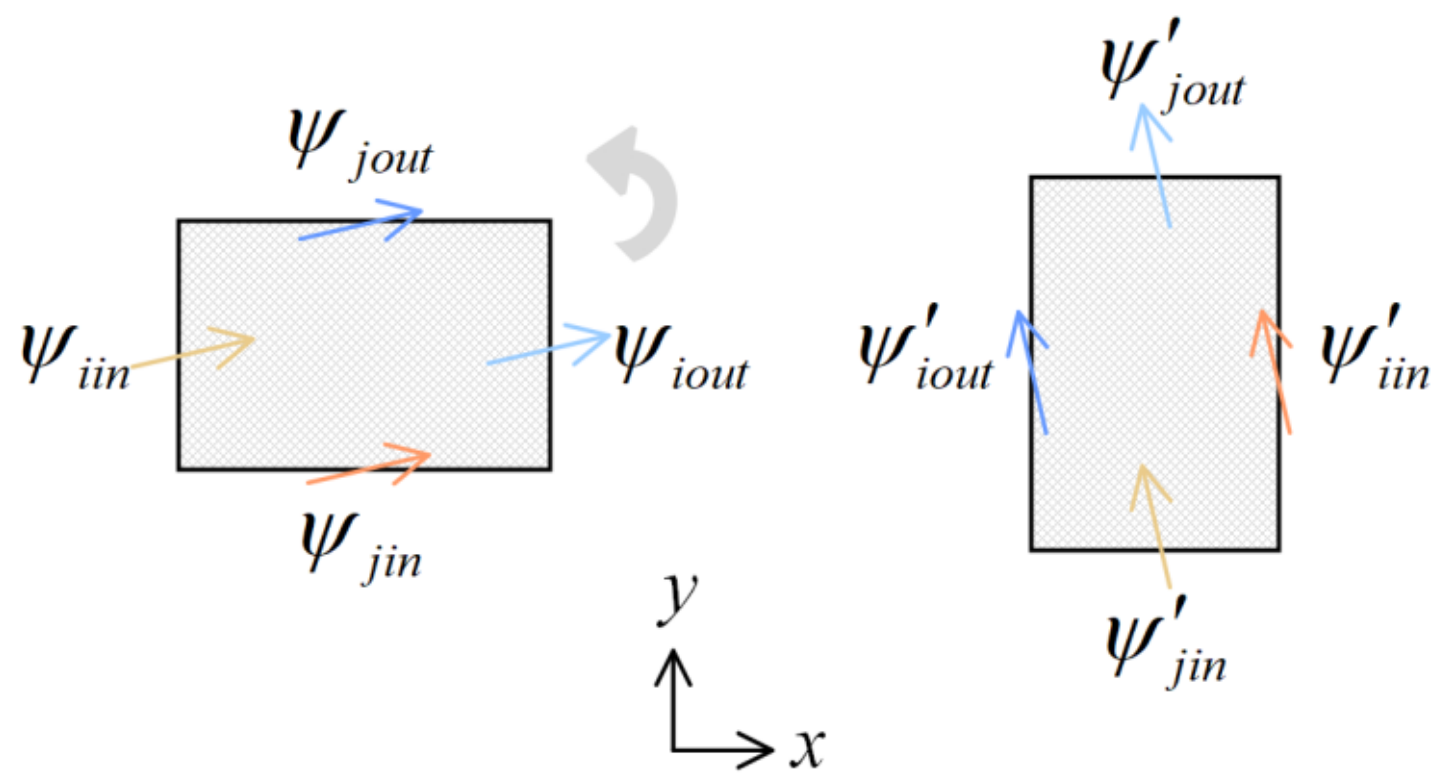

Fig. 2. Diagram of permutation-equivariant property

Neural networks cannot inherently guarantee that inputs and outputs satisfy this property. Adopting data augmentation approach, we expand the training dataset by applying rotational transformations to both inputs and outputs. The network inputs can be categorized into $\mathbf{u}_0 = \{\bar{Q}/\Sigma_t \bar{\psi}\}$ and the corresponding quantities for each dimension as

$$\mathbf{u}_x = \left\{ \frac{\Sigma_t \Delta x}{|\mu|}, \frac{\psi_{iin}}{\bar{\psi}}, \frac{\psi_{iout}}{\bar{\psi}}, \frac{Q_x}{\bar{Q}} \right\}$$
$$\mathbf{u}_y = \left\{ \frac{\Sigma_t \Delta y}{|\eta|}, \frac{\psi_{jin}}{\bar{\psi}}, \frac{\psi_{jout}}{\bar{\psi}}, \frac{Q_y}{\bar{Q}} \right\} \tag{38}$$
$$\mathbf{u}_z = \left\{ \frac{\Sigma_t \Delta z}{|\xi|}, \frac{\psi_{kin}}{\bar{\psi}}, \frac{\psi_{kout}}{\bar{\psi}}, \frac{Q_z}{\bar{Q}} \right\}$$

By transforming the order of the inputs, the output quantities exhibit the following corresponding relationships:

$$\begin{aligned}
f_{NN}\left(\mathbf{u}_x,\mathbf{u}_y,\mathbf{u}_z,\mathbf{u}_0\right) &= \left\{ f_{iout}, f_{jout}, f_{kout} \right\} \\
f_{NN}\left(\mathbf{u}_y,\mathbf{u}_z,\mathbf{u}_x,\mathbf{u}_0\right) &= \left\{ f_{jout}, f_{kout}, f_{iout} \right\} \\
f_{NN}\left(\mathbf{u}_z,\mathbf{u}_x,\mathbf{u}_y,\mathbf{u}_0\right) &= \left\{ f_{kout}, f_{iout}, f_{jout} \right\} \\
f_{NN}\left(\mathbf{u}_x,\mathbf{u}_z,\mathbf{u}_y,\mathbf{u}_0\right) &= \left\{ f_{iout}, f_{kout}, f_{jout} \right\} \\
f_{NN}\left(\mathbf{u}_y,\mathbf{u}_x,\mathbf{u}_z,\mathbf{u}_0\right) &= \left\{ f_{jout}, f_{iout}, f_{kout} \right\} \\
f_{NN}\left(\mathbf{u}_z,\mathbf{u}_y,\mathbf{u}_x,\mathbf{u}_0\right) &= \left\{ f_{kout}, f_{jout}, f_{iout} \right\}
\end{aligned} \tag{39}$$

A set of training samples is expanded into six groups, and corresponding loss terms are added. To prevent overfitting, an L2 regularization term is also added. Finally, the loss function for pcNN is

$$\begin{aligned}
loss = {} & w_1\left(loss_{out}^{xyz} + loss_{out}^{yzx} + loss_{out}^{zxy} + loss_{out}^{xzy} + loss_{out}^{yxz} + loss_{out}^{zyx}\right) + \\
& w_2 loss_{phy} + w_3 loss_{L2}
\end{aligned} \tag{40}$$

where $w$ is the weighting coefficient for each loss item.

# 5. pcNN-$S_N$ coupling algorithm

## 5.1 Training data generation

By simulating several typical transport problems, we can obtain high- and low-accuracy angular fluxes while generating local features and correction factors for coarse-grid solutions. We generate and record the training data during global fine-grid simulations. By projecting the fine-grid physical quantities onto the coarse grid, we obtain the equivalent physical parameters for that local coarse grid, which can then be used to compute a local low-accuracy angular flux.

The fine-to-coarse ratio is manually set. Taking a two-dimensional case in Fig. 3 as an example, the coarsening ratio for each dimension is set to 2, meaning that 4 fine grids correspond to 1 coarse grid. Following the principle of volume conservation, the equivalent source strength and incoming angular fluxes are calculated as

$$\tilde{Q} = \frac{\sum_{\Delta V_n \in D} \bar{Q}_n \Delta V_n}{\sum_{\Delta V_n \in D} \Delta V_n} \tag{41}$$

$$\tilde{\psi}_{iin} = \frac{\sum_{\Delta A_n \in \partial D} \psi_{iin,n} \Delta A_n}{\sum_{\Delta A_n \in \partial D} \Delta A_n} \tag{42}$$

where $D$ and $\partial D$ denote the computational domain and its boundary of the coarse grid, respectively, $\Delta V_n$ and $\Delta A_n$ denote the volume and area of the $n$-th fine grid. Similarly, projecting the high-accuracy solutions of the fine grids onto the coarse grid generates the reference solution as

$$\bar{\psi}^{high} = \frac{\sum_{\Delta V_n \in D} \bar{\psi}_n \Delta V_n}{\sum_{\Delta V_n \in D} \Delta V_n} \tag{43}$$

$$\psi_{iout}^{high} = \frac{\sum_{\Delta A_n \in \partial D} \psi_{iout,n} \Delta A_n}{\sum_{\Delta A_n \in \partial D} \Delta A_n} \tag{44}$$

After performing local angular flux calculations based on coarse-grid equivalent parameters, we can get the correction factors for the low-accuracy solution, as well as their local physical features. Upon completing each simulation of a coarse grid, a set of input-output training data is obtained.

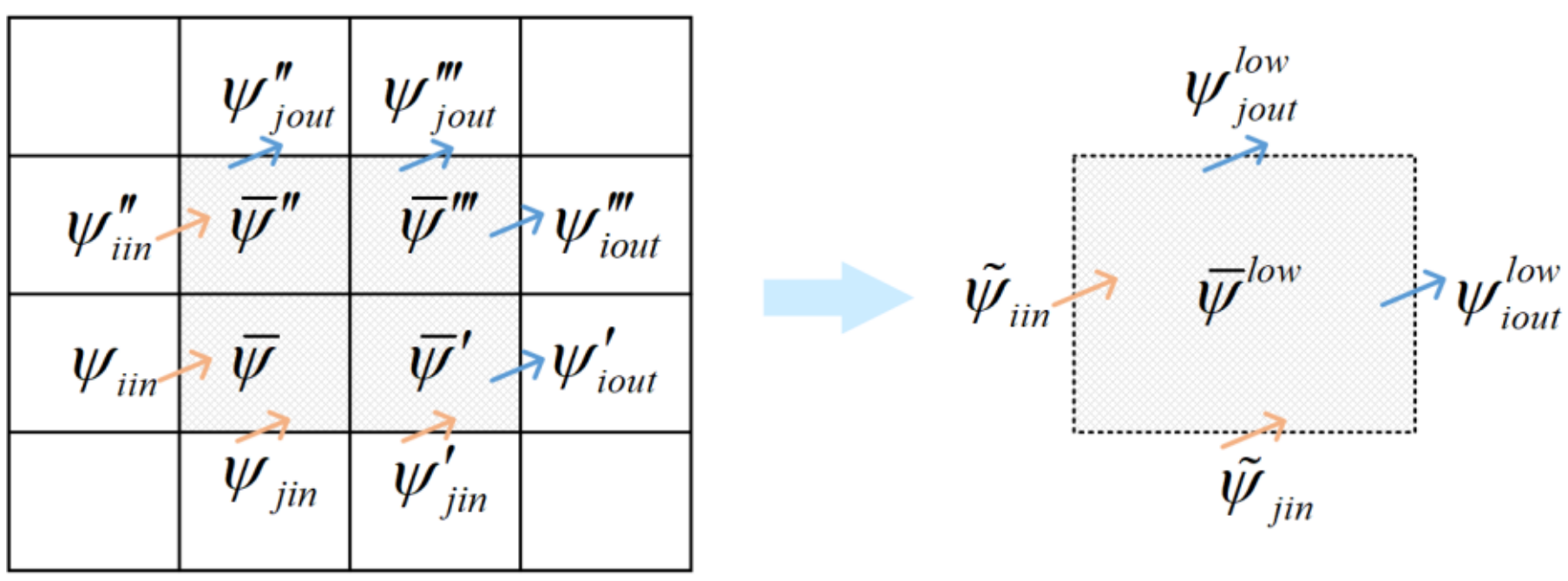

Fig. 3. Generation of local coarse-grid equivalent parameters and low-accuracy solution

### 5.2 Coupling framework

After constructing the training dataset, the pcNN model is trained to predict correction factors for angular fluxes of local coarse grids. Our model is built using the PyTorch platform. The network architecture employs a fully connected neural network with two hidden layers, each containing 8 neurons. The ReLU activation function is used, and the Adam optimizer is utilized for training networks. Upon meeting loss criteria or reaching the maximum iteration count, the weights and biases of network model are recorded for later use by the $S_N$ transport code, as shown in Fig.4.

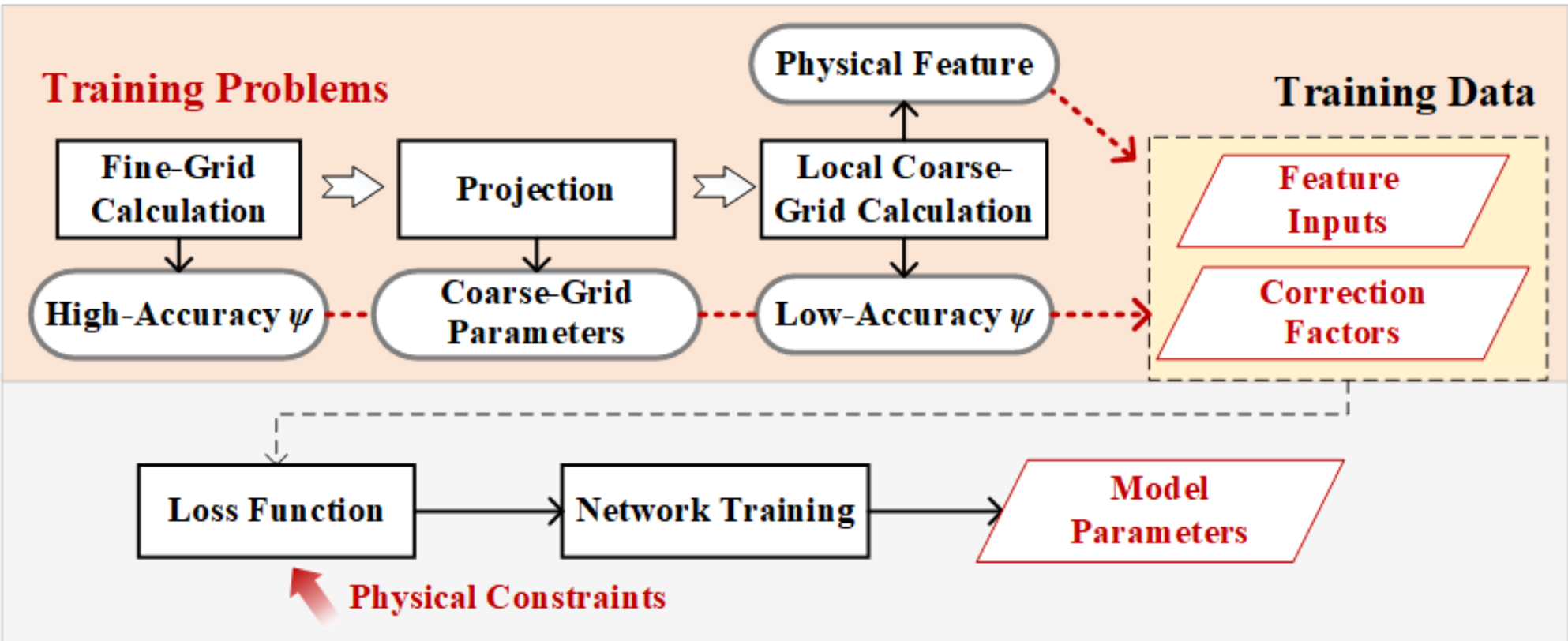

Fig. 4. Construction of the pcNN model

Unlike existing studies for global physical fields, our machine learning model is specifically designed for local numerical solutions. An internal coupling strategy is employed to integrate our pcNN model with $S_N$ calculations. The traditional $S_N$ process and pcNN-$S_N$ calculation processes are shown in Figs. 5 and 6, respectively. $S_N$ transport sweeps require cycles of discrete directions and spatial grids, solving each grid sequentially based on dependency relationships. After solving a specific direction and grid, we construct local features of the angular flux solution and feed them into the neural network. Based on trained network parameters, correction factors are generated and online correction is performed. The corrected outgoing angular flux solution is then passed to the downwind grid to continue $S_N$ sweep calculations.

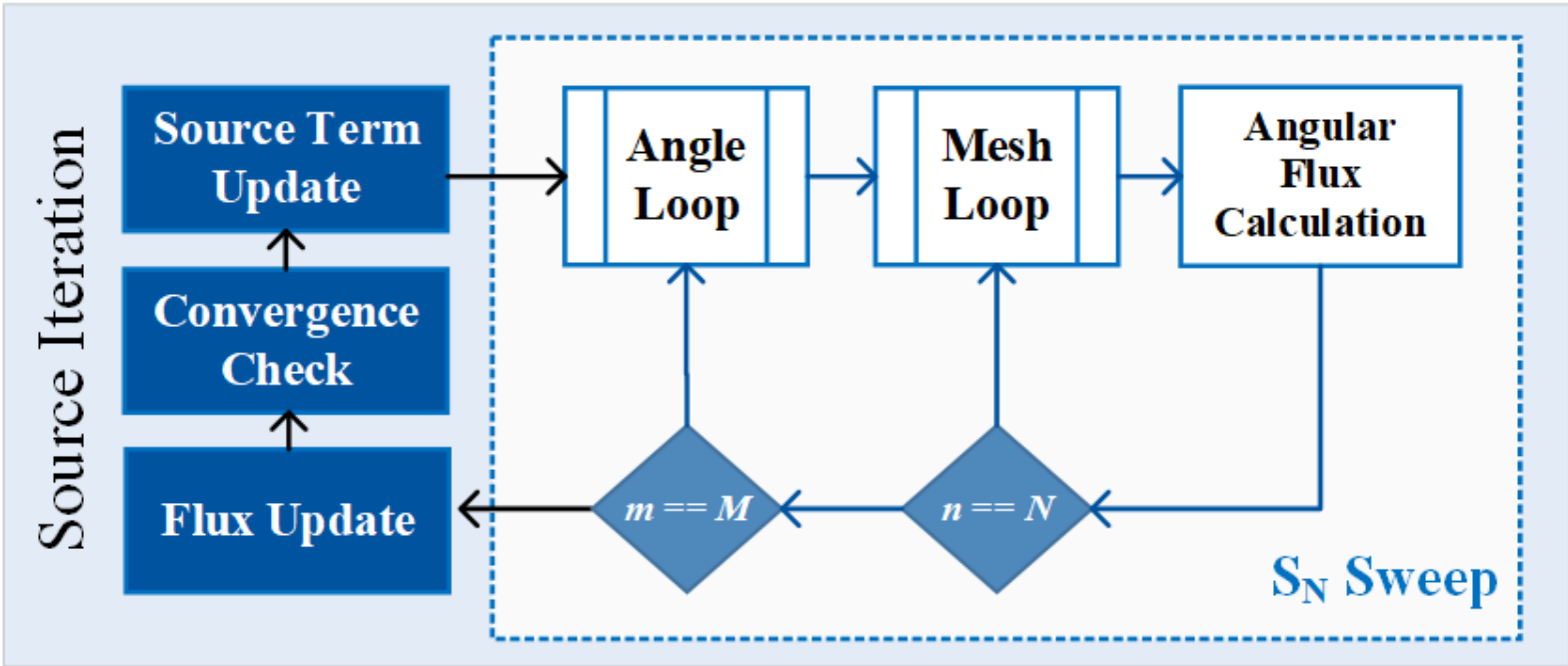

Fig. 5. Calculation process of traditional $S_N$ algorithm

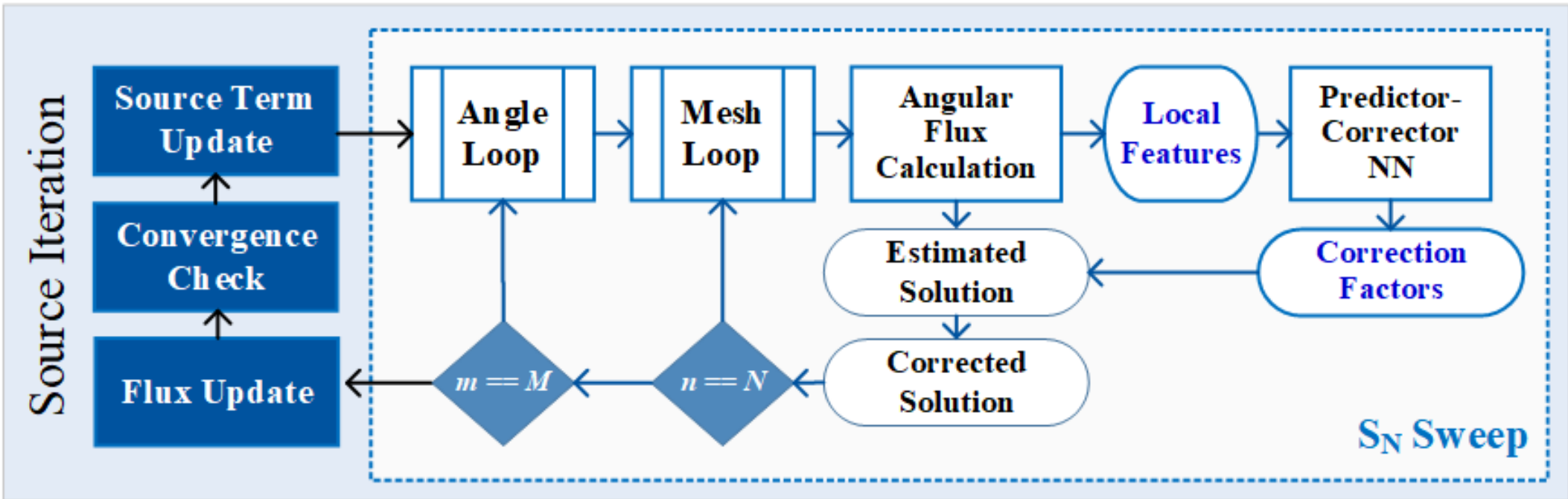

Fig. 6. Calculation process of pcNN-$S_N$ coupling algorithm

## 6. Numerical verification

Our pcNN-$S_N$ coupling algorithm is tested for the 1-D and 3-D Cartesian geometries. The training problems are simulated to acquire data and network models are constructed. For the test problems, we vary physical and computational parameters such as material cross section, grid size, computational domain, etc., to verify the correctness of the coupling algorithm and the reusability of the model.

### 6.1 1-D single-medium inflow problem

For the training problems, the left boundary is a given inflow boundary, while the right boundary is a vacuum boundary. The total cross section and scattering cross section are 1 $cm^{-1}$ and 0.9 $cm^{-1}$, respectively. An $S_8$-order quadrature set is used. Adopting a grid size of 0.0625 cm, simulations are conducted for four cases with computational domains of 3 cm, 6 cm, 9 cm, and 12 cm, respectively. The resulting flux distributions are shown in Fig. 7. The fine-grid data are projected onto coarse grids of 1 cm, 2 cm, 3 cm, and 4 cm, respectively, to form a training dataset.

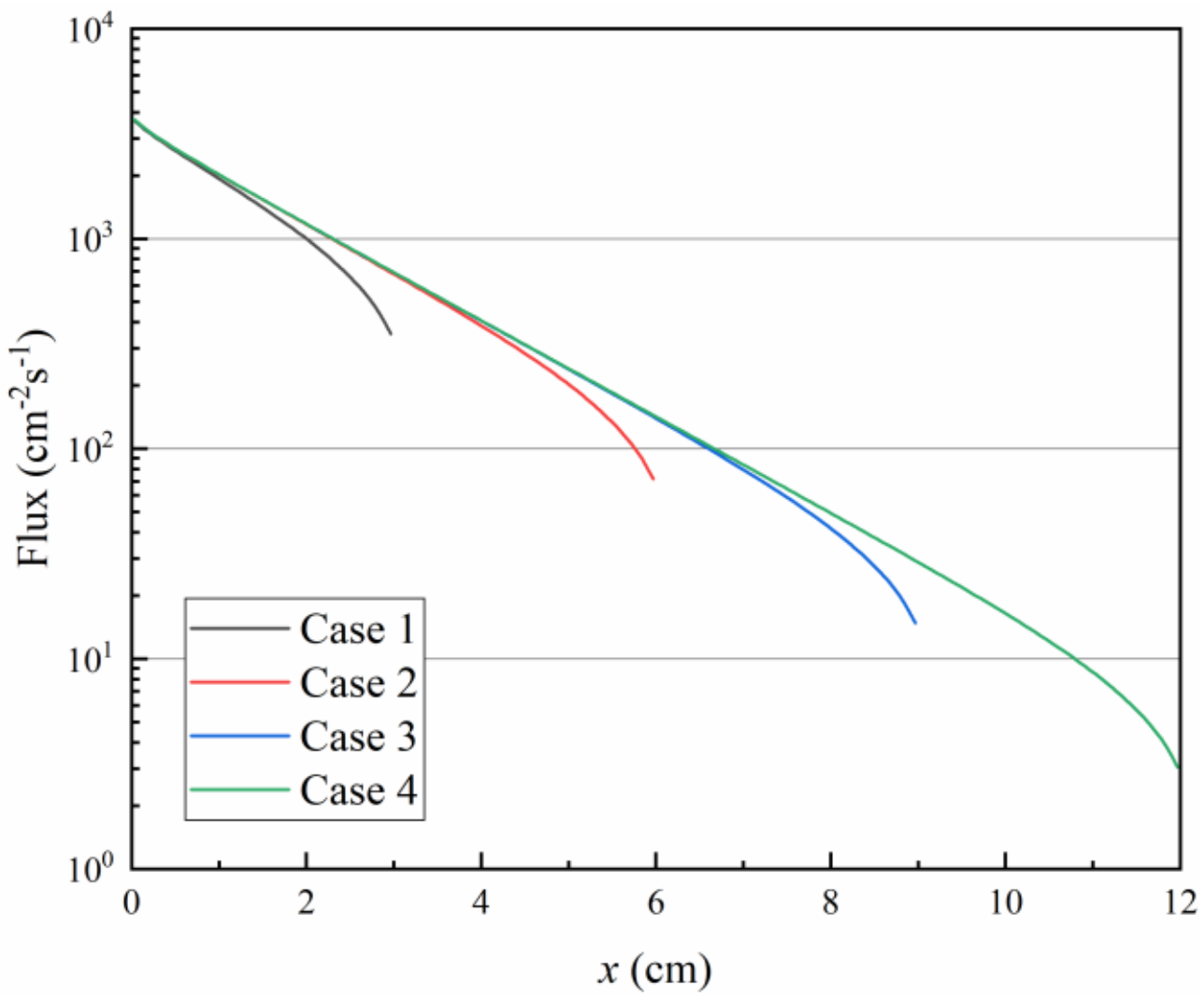

Fig. 7. Fine-grid flux distributions for the 1-D inflow training problems

The test problem varies the left boundary inflow, geometry, cross sections, and mesh size. The computational domain is 20 cm. The total cross section is 1.2 cm⁻¹. The scattering cross sections are set to 0 and 1.0 cm⁻¹, respectively. The simulation employs a grid spacing of 2.5 cm, and flux results of coarse-grid $S_N$ and pcNN-$S_N$ calculations are compared with fine-grid $S_N$ reference, as shown in Fig. 8. Due to the large grid spacing, the traditional coarse-grid $S_N$ difference results deviate significantly from the reference. The pcNN-$S_N$ calculation results show good agreement with reference.

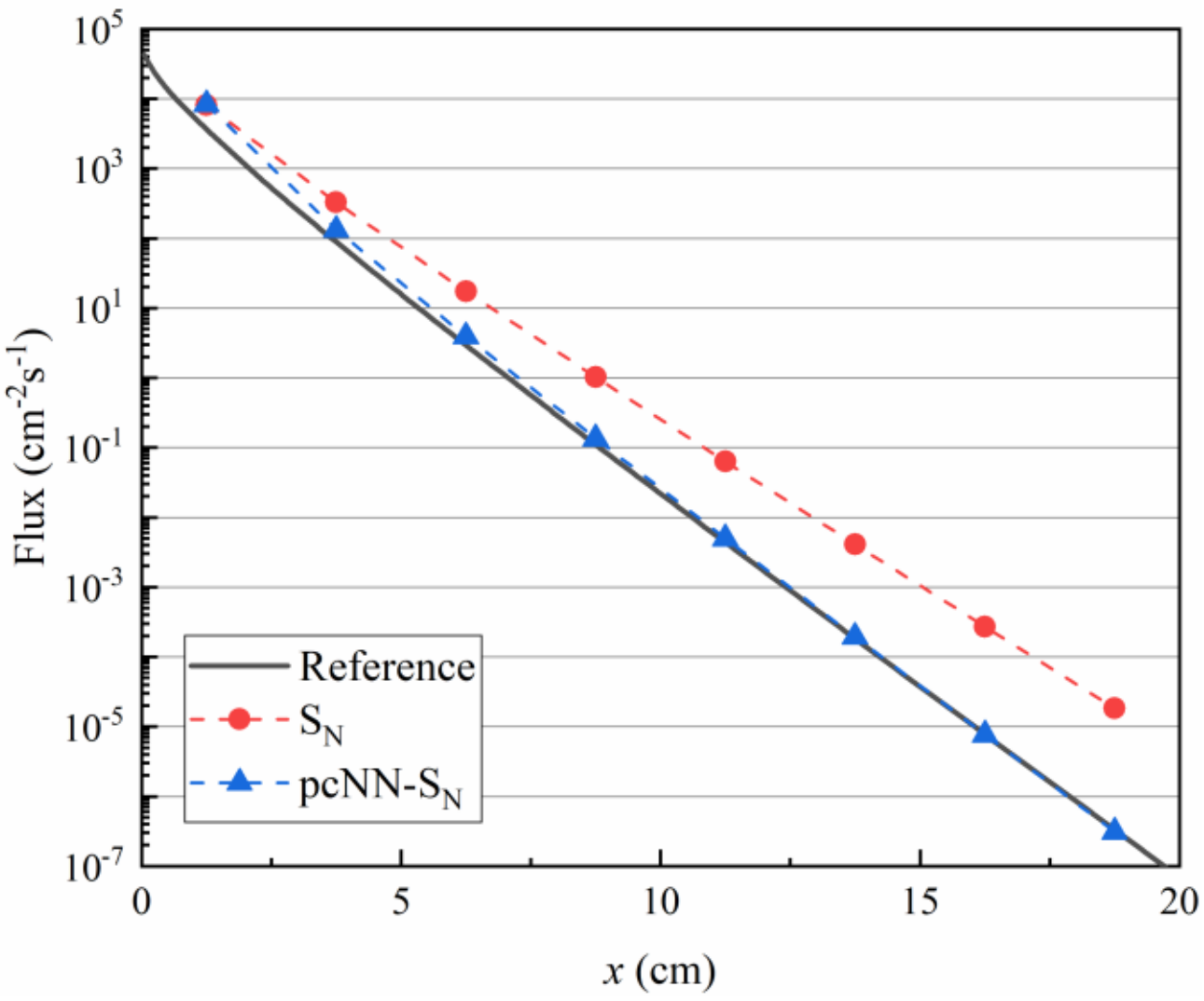

(a) without scattering

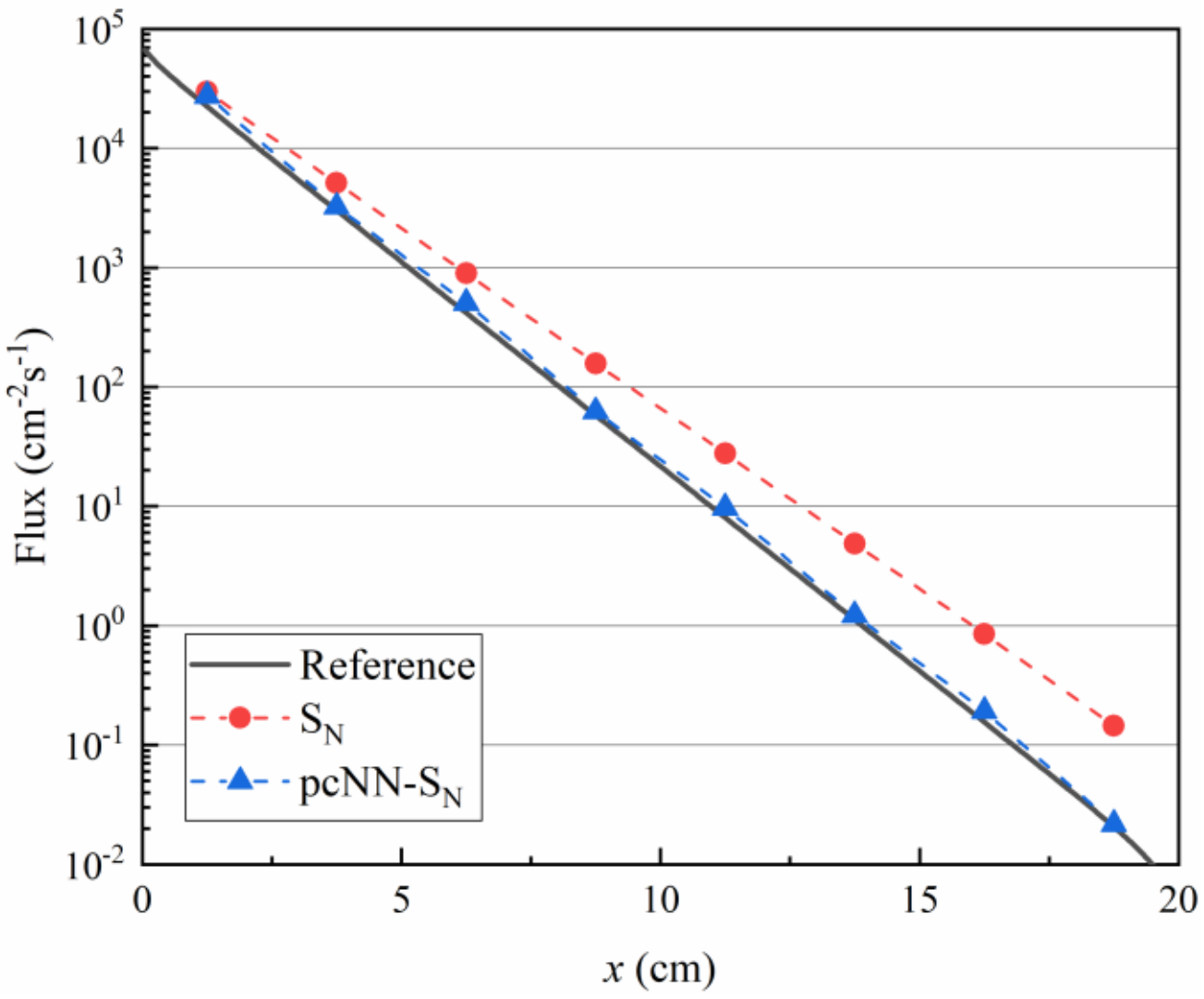

(b) with scattering

Fig. 8. Flux results for the 1-D inflow test problem

The computational parameters of the test problem differ from the training problems. However, after parameter dimensionless normalization, the inputs for the test problem fall within the parameter range of the training dataset. Based on local physical features and internal coupling strategies, our neural network model demonstrates good reusability under various conditions. As long as the model can capture the correspondence between high- and low-accuracy solutions within the parameter range, it can effectively correct coarse-grid solutions.

### 6.2 1-D single-medium external source problem

The left boundary of the training problems is a reflective boundary, while the right boundary is a vacuum boundary. The total and scattering cross sections are 1 cm$^{-1}$ and 0.9 cm$^{-1}$ respectively. An external source is placed in the left half-space. Using a grid size of 0.0625 cm, simulations are conducted with computational domains of 10 cm, 20 cm, 30 cm, and 40 cm, respectively, whose fluxes are shown in Fig. 9. The simulation data are projected onto coarse grids of 1 cm, 2 cm, 3 cm, and 4 cm, respectively, to form a training dataset.

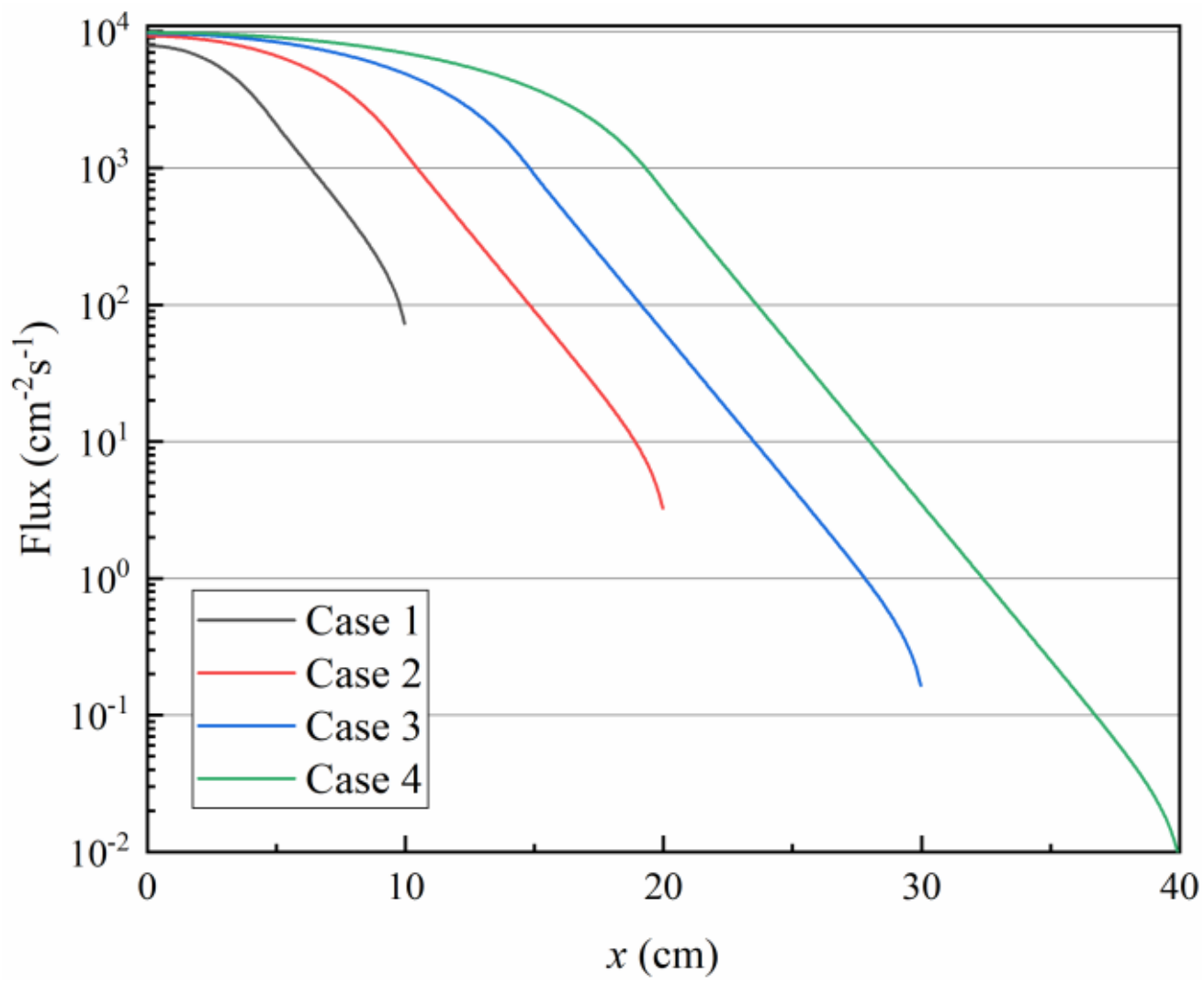

Fig. 9. Fine-grid flux distributions for the 1-D external source training problems

The test problem alters boundary conditions, geometry, source regions, cross sections, and mesh size. All boundary conditions are vacuum boundaries. The computational domain size is 60 cm. Two non-connected source regions are located within the intervals [10, 20] cm and [40, 50] cm, respectively. The total and scattering cross sections are 0.9 cm$^{-1}$ and 0.1 cm$^{-1}$ respectively. Using a 2.5 cm grid, numerical results are shown in Fig. 10. Within two source regions, the flux distribution exhibits a flat profile, with both coarse-grid $S_N$ and pcNN-$S_N$ calculations showing good agreement with the reference. In regions of strong flux attenuation, the coarse-grid $S_N$ results deviate significantly from the fine-grid solution, while the coupling algorithm results still show good agreement. The global flux field of test problems differs markedly from that of training problems. Our method is based on local solution correction, so its generalization ability is not limited in such scenarios.

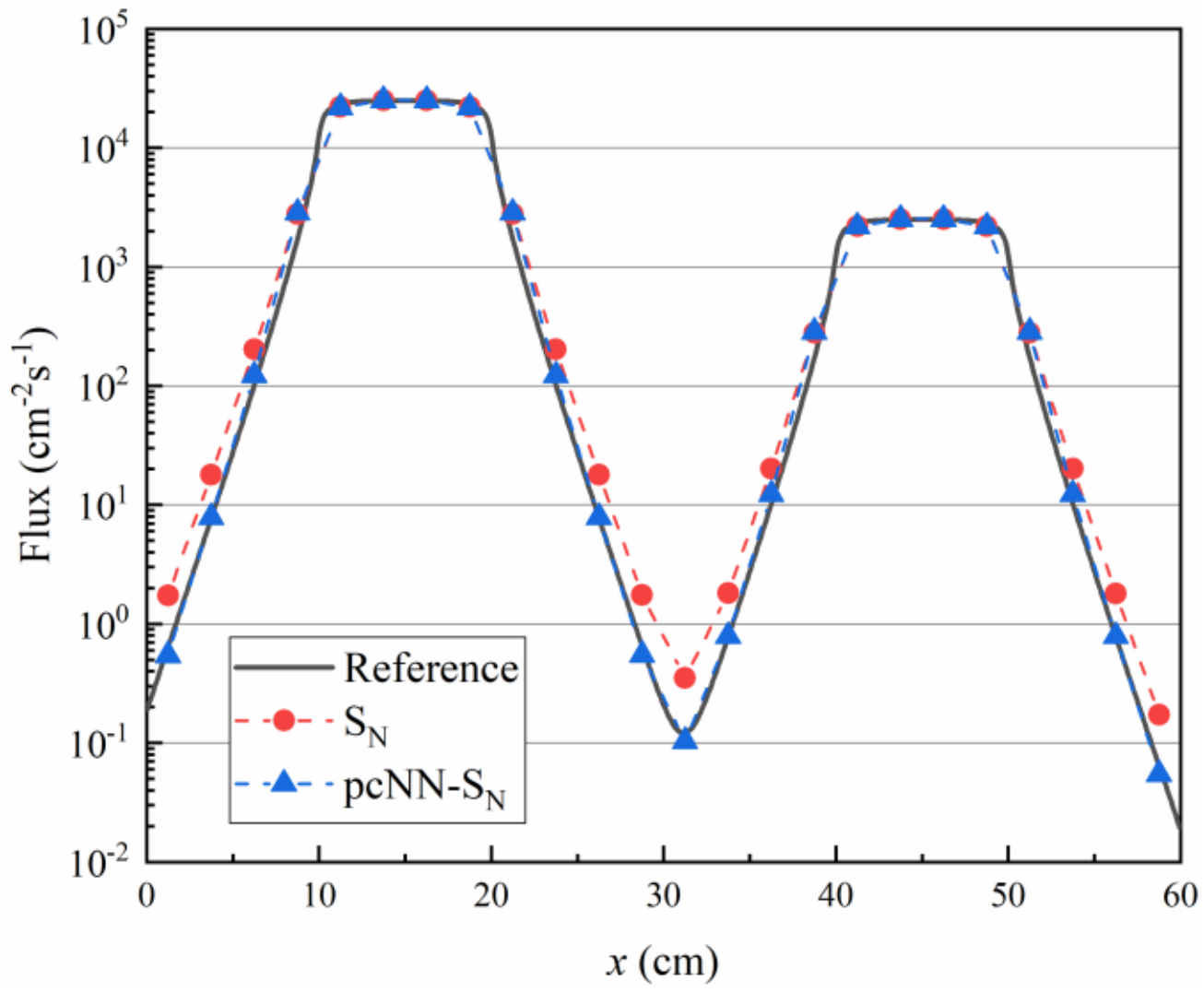

Fig. 10. Flux results for the 1-D external source test problem

### 6.3 1-D two-medium external source problem

Based on the training problem setup described in Section 6.2, we add training data for other grid sizes and test our algorithm on a two-medium problem. The geometric structure of the test problem is shown in Fig. 11. The left and right boundaries are defined as reflective and vacuum boundary conditions, respectively. Material region 1 covers the range from 0 to 20 cm, with a total cross section of 1.5 cm$^{-1}$ and a scattering cross section of 0.5 cm$^{-1}$. Material region 2 covers the range from 20 to 40 cm, with a total cross section of 0.5 cm$^{-1}$ and a scattering cross section of 0.4 cm$^{-1}$. The external source is located from 0 to 10 cm.

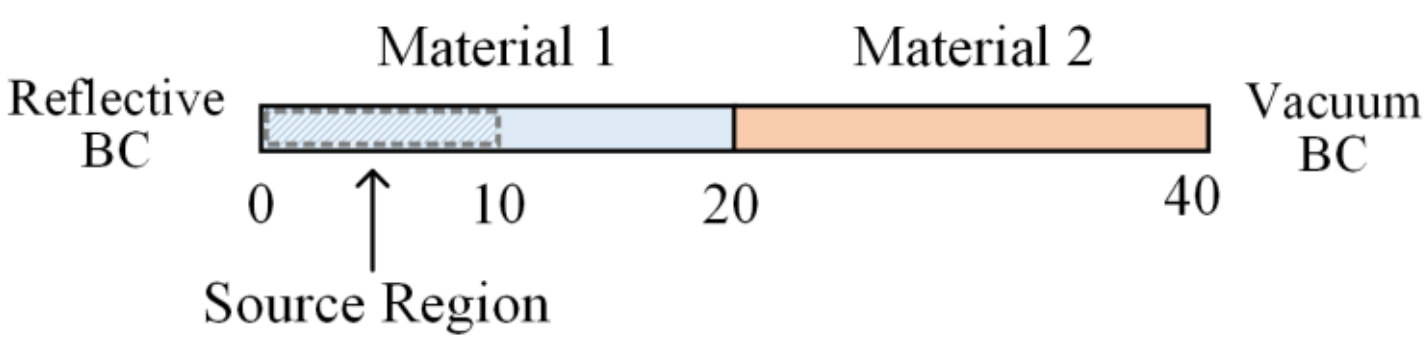

Fig. 11. Geometric model of the 1-D two-medium problem

The results using 2-cm grids are shown in Fig. 12. Within a range of10 to 20 cm, the attenuation of the flux is quite significant. Our new method yields significantly better results than traditional method on coarse grids. The pcNN-$S_N$ algorithm relaxes the constraints on grid spacing imposed by difference schemes.

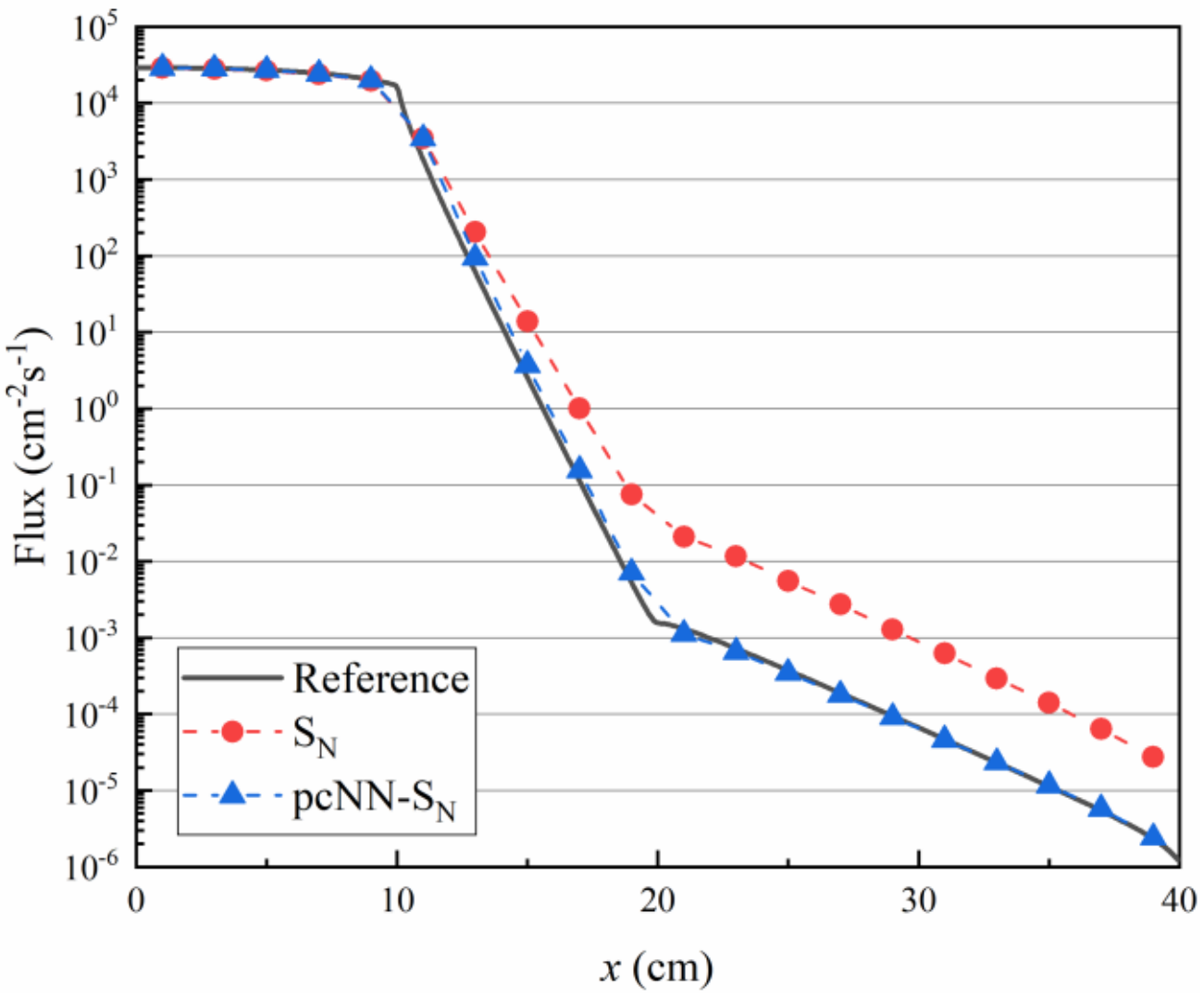

Fig. 12. Flux results for the 1-D two-medium problem

### 6.4 3-D single-medium problem without scattering

The computational domain for the training problem is a cubic region with an edge length of 8 cm. Reflective boundary conditions are applied to the -x, -y, and -z boundaries, while vacuum boundary conditions are applied to the +x, +y, and +z boundaries. The total cross section is 2.2 cm$^{-1}$. We set the external source throughout the entire system as

$$q(x,y,z) = c_0 e^{-c_1 x} e^{-c_1 y} e^{-c_1 z} \tag{45}$$

where $c_0$ and $c_1$ denote given constant coefficients. We project the fine-grid simulation data from 0.125-cm grids onto coarse grids of 1 cm, and generate the training dataset.

After the network model is constructed, we calculate this problem using the same coarse grids as in model training. Using the fine-grid results as a reference, two error norms are calculated as

$$\mathrm{E}_{\mathrm{RMS}} = \sqrt{\frac{1}{N}\sum_{n=1}^{N}\left(\frac{\phi_n^{calc} - \phi_n^{ref}}{\phi_n^{ref}}\right)^2} \tag{46}$$

$$\mathrm{E}_{\mathrm{inf}} = \max_n \left\{ \left| \frac{\phi_n^{calc} - \phi_n^{ref}}{\phi_n^{ref}} \right| \right\} \tag{47}$$

where $\phi_n^{calc}$ and $\phi_n^{ref}$ are the calculated and reference values of scalar fluxes for the $n$-th grid, respectively. $\mathrm{E}_{\mathrm{RMS}}$ norms for the coarse-grid $S_N$ and pcNN-$S_N$ calculations are 2.01% and 0.16%, respectively, and $\mathrm{E}_{\mathrm{inf}}$ norms are 7.16% and 0.53%, respectively. For this training problem, our network model effectively captures the patterns of error variation.

In the test problem, the geometry, cross section, and grid size are changed. The geometric length of each dimension is 15 cm. The total cross section is 1.6 cm$^{-1}$, and the grids of 1.25 cm are used. The $S_N$ results obtained using a grid of 0.15625 cm are used as a reference. $E_{RMS}$ norms for the $S_N$ and pcNN-$S_N$ calculations are 2.78% and 0.54%, respectively, with $E_{inf}$ of 10.56% and 1.37%, respectively. Fig. 13 shows their flux results and the absolute values of relative error distributions. Using the same grid, our new algorithm yields more accurate results.

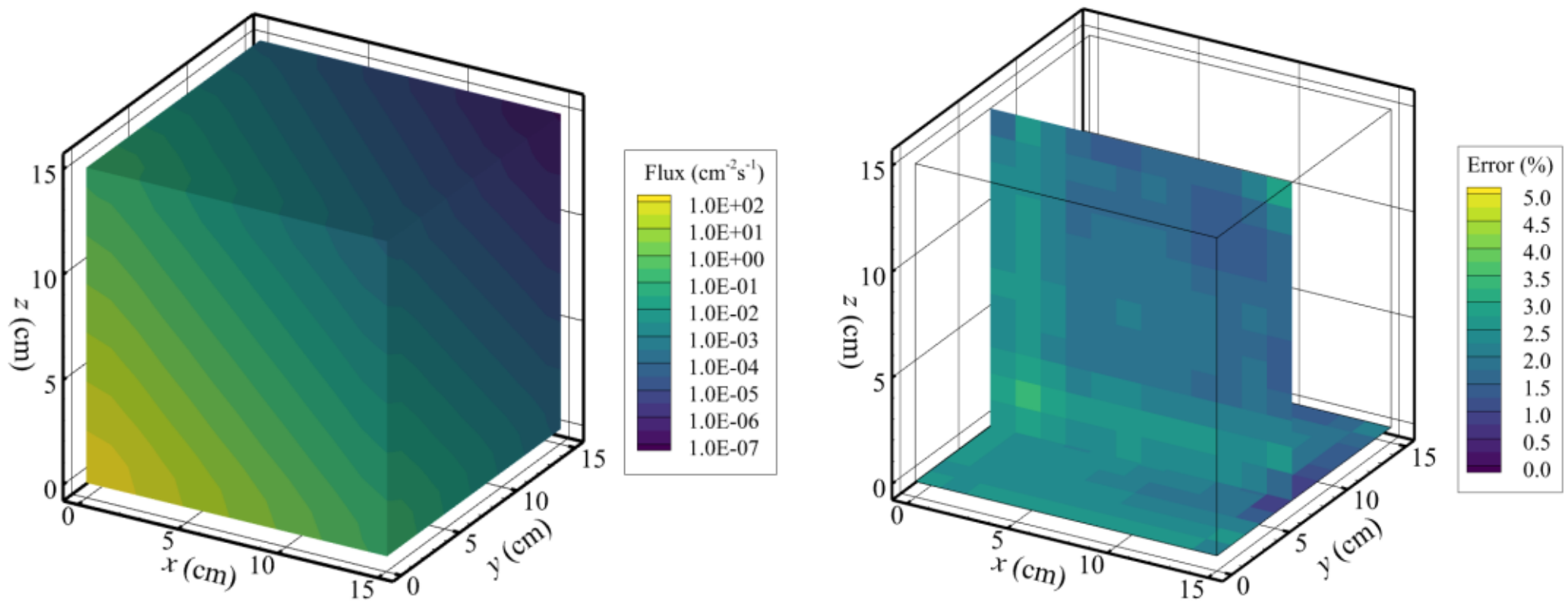

(a) $S_N$

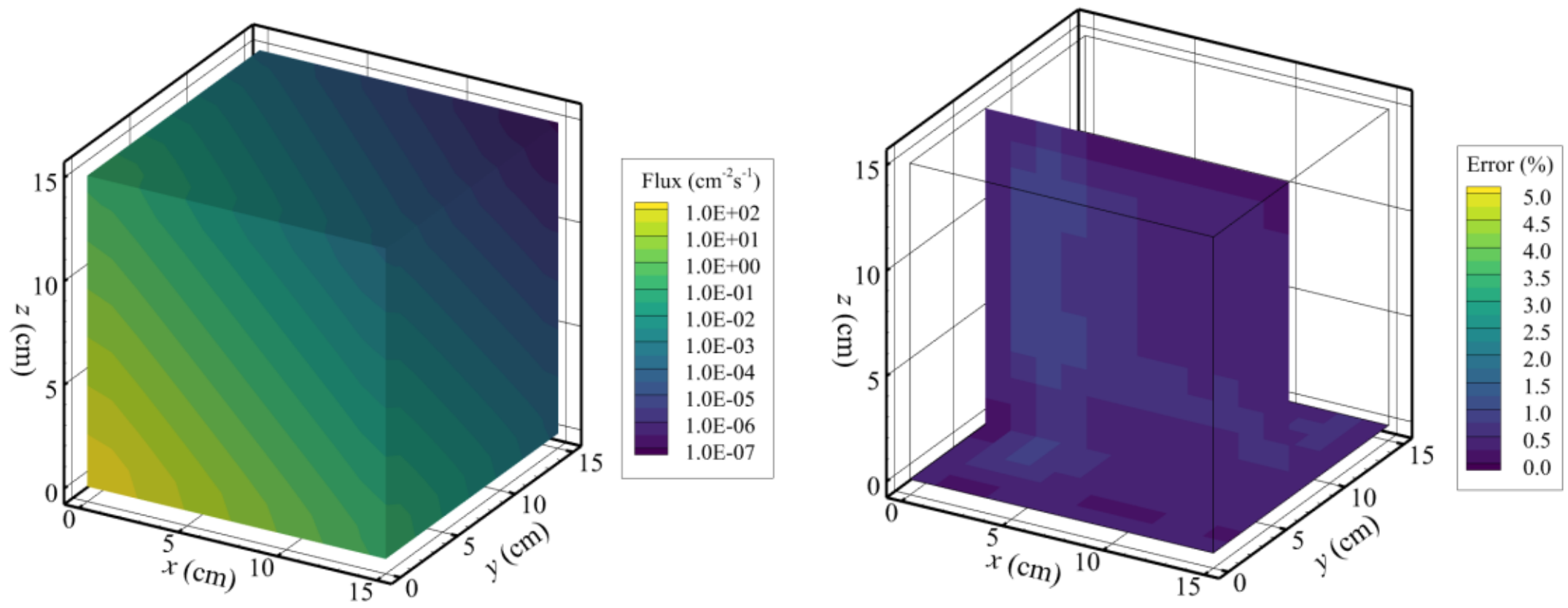

(b) pcNN-$S_N$

Fig. 13. Results for 3-D single-medium test problem without scattering

### 6.5 3-D single-medium problem with scattering

The geometric model of the training problem is an 8 cm cube. The total and scattering cross sections are 2.2 cm$^{-1}$ and 0.6 cm$^{-1}$ respectively. The external source distribution is identical to that in Sec. 6.4. The fine grid size is 0.125 cm, generating training data corresponding to 1-cm coarse grids. For this training case, $E_{RMS}$ norms for the coarse-grid $S_N$ and pcNN-$S_N$ calculations are 2.05% and 0.46%, respectively, and $E_{inf}$ norms are 7.77% and 1.50%, respectively.

In the test problem, the geometric length is 16 cm. The total cross section and scattering cross section are 2.5 $cm^{-1}$ and 0.5 $cm^{-1}$, respectively. Take the $S_N$ results of 0.1-cm grids as a reference. We perform traditional $S_N$ calculations using 0.8-cm and 0.4-cm grids, while the pcNN-$S_N$ algorithm only uses a grid of 0.8 cm. We list the error norms in Table 1. The absolute values of relative errors in the flux distribution for each case are shown in the Fig. 14. And the accuracy of the pcNN-$S_N$ algorithm is better. The $S_N$ calculation on 0.4 cm grids take 9.5 seconds, while the pcNN-$S_N$ calculation on 0.8 cm grids take 3.3 seconds. Our algorithm achieves lower error and shorter computation time. The cost of network evaluation is lower than the $S_N$ calculation cost resulting from grid refinement. Compared to $S_N$ calculations on a refined grid, the coupling algorithm reduces computation time by 65%.

Table 1. Error norms for 3-D single-medium test problem with scattering

| Algorithm | Mesh size | $E_{RMS}$ | $E_{inf}$ |
|---|---|---|---|
| $S_N$ | 0.8 cm | 0.97% | 5.24% |
| $S_N$ | 0.4 cm | 0.38% | 2.93% |
| pcNN-$S_N$ | 0.8 cm | 0.23% | 0.96% |

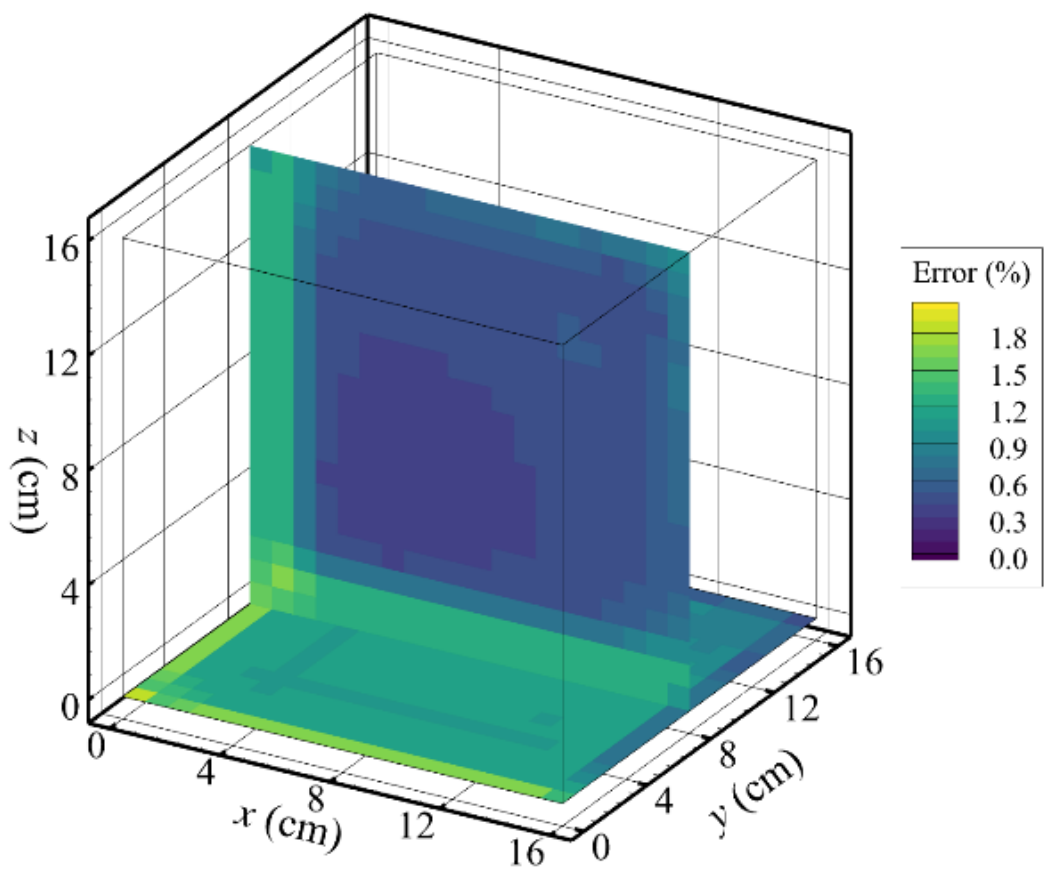

(a) $S_N$ using 0.8-cm grids

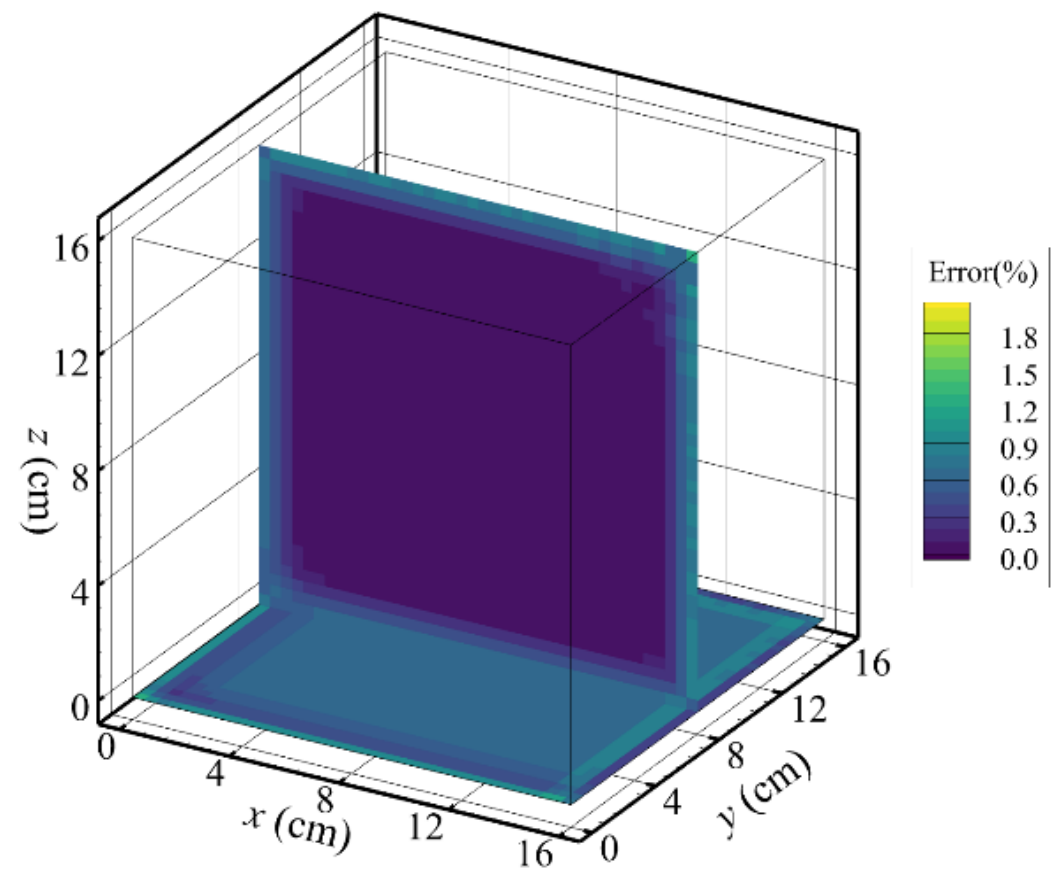

(b) $S_N$ using 0.4-cm grids

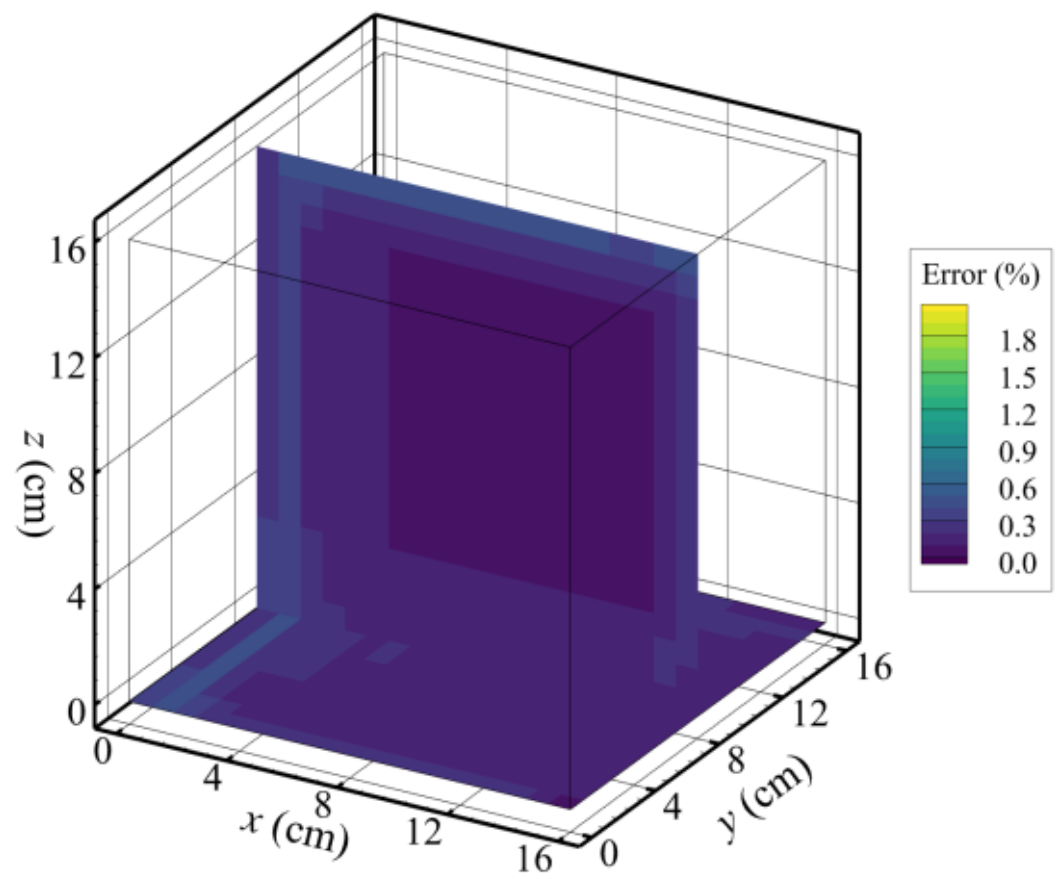

(c) pcNN-$S_N$ using 0.8-cm grids

Fig. 14. Error results for 3-D single-medium test problem with scattering

### 6.6 Time-dependent two-medium problem with variable cross sections

For time-dependent problems, the impact of time discretization is temporarily disregarded, and the time step for both coarse- and fine-grid $S_N$ calculations is identical. The training problem is a simple box-in-box cube model with an edge length of 6.4 cm, containing two material regions. Material region 1 is located within a cubic region ranging from 0 to 3.2, and its exterior is surrounded by material region 2. At the initial moment, the total cross section and scattering cross section of material 1 are 0.8 cm$^{-1}$ and 0.3 cm$^{-1}$ respectively, while those of material 2 are 1.2 cm$^{-1}$ and 0.25 cm$^{-1}$ respectively. The cross sections of both materials changes over time, while their scattering ratio remains constant. Initial flux is set to 0, and $v\Delta t$ is set to 1, and 15 time steps are simulated. The fine grid size is 0.1 cm, and the training dataset are generated by projecting fine-grid data onto a coarse grid of 0.8 cm.

The test problem alters geometric dimensions and modify the temporal evolution of material cross sections. The length of the geometric model is 16 cm. Material region 1 occupies the cubic region from 0 to 6.4 cm, while the remaining region belongs to material region 2. 20 time steps are simulated. The variation of the total cross sections over time is shown in Fig. 15.

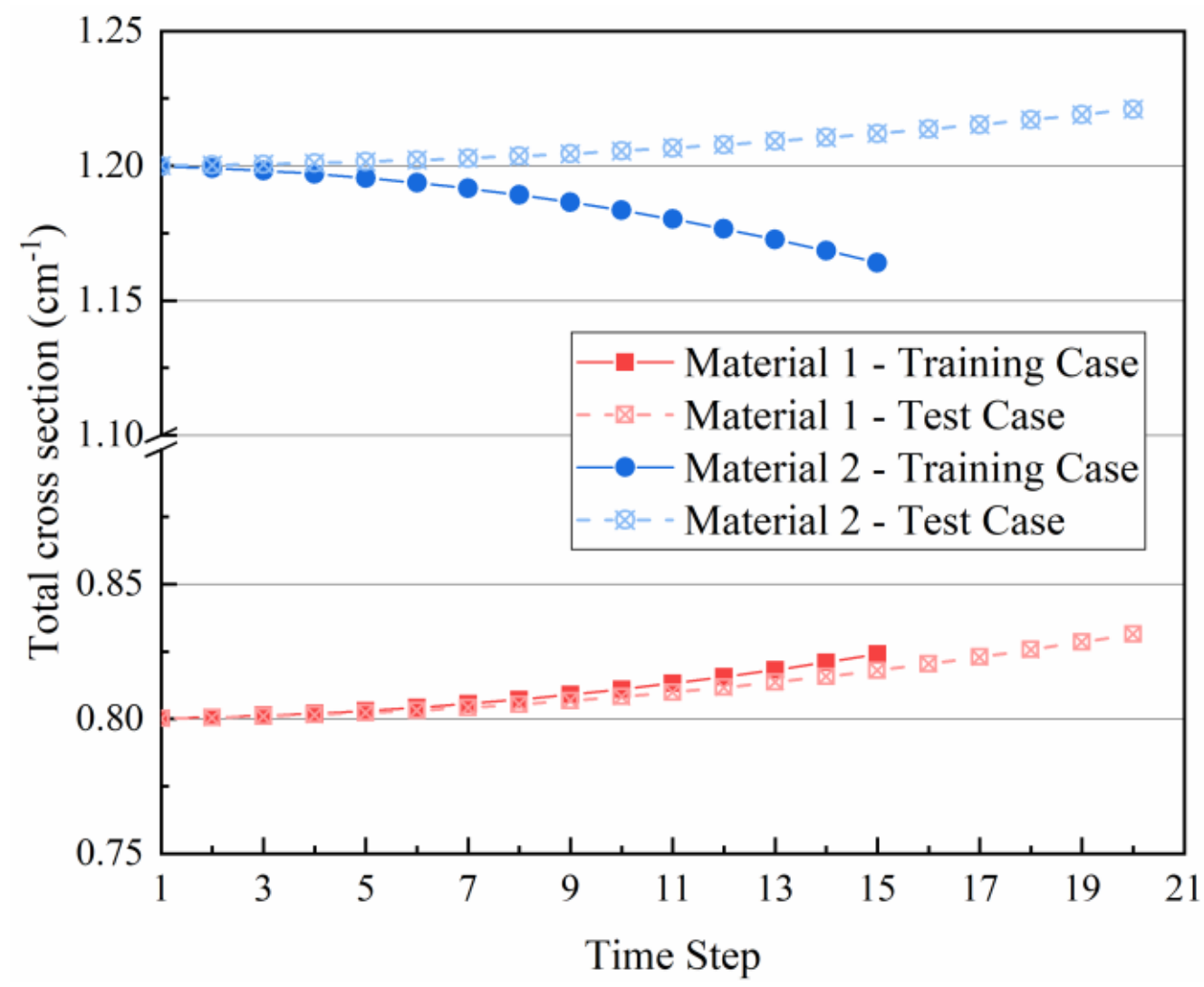

Fig. 15. Evolution of total cross sections for the 3-D time-dependent problems

Take the $S_N$ results with grids of 0.1 cm as a reference. The traditional $S_N$ simulations use 0.4- and 0.8-cm grids, and the pcNN-$S_N$ algorithm uses 0.8-cm grids. Results at the final time step are shown in Fig. 16. For all time steps, traditional $S_N$ simulations based on 0.8 cm and 0.4 cm grids produce RMS relative errors of 1.70% and 0.94%, respectively. The pcNN-$S_N$ algorithm achieves an $E_{RMS}$ norm of 0.64%, which is lower than the $S_N$ results obtained using the same grid and refined grid. In the test case, the variation ranges of cross sections of material 2 exceed the training data range, yet the neural network model remains effective in the coupling calculations.

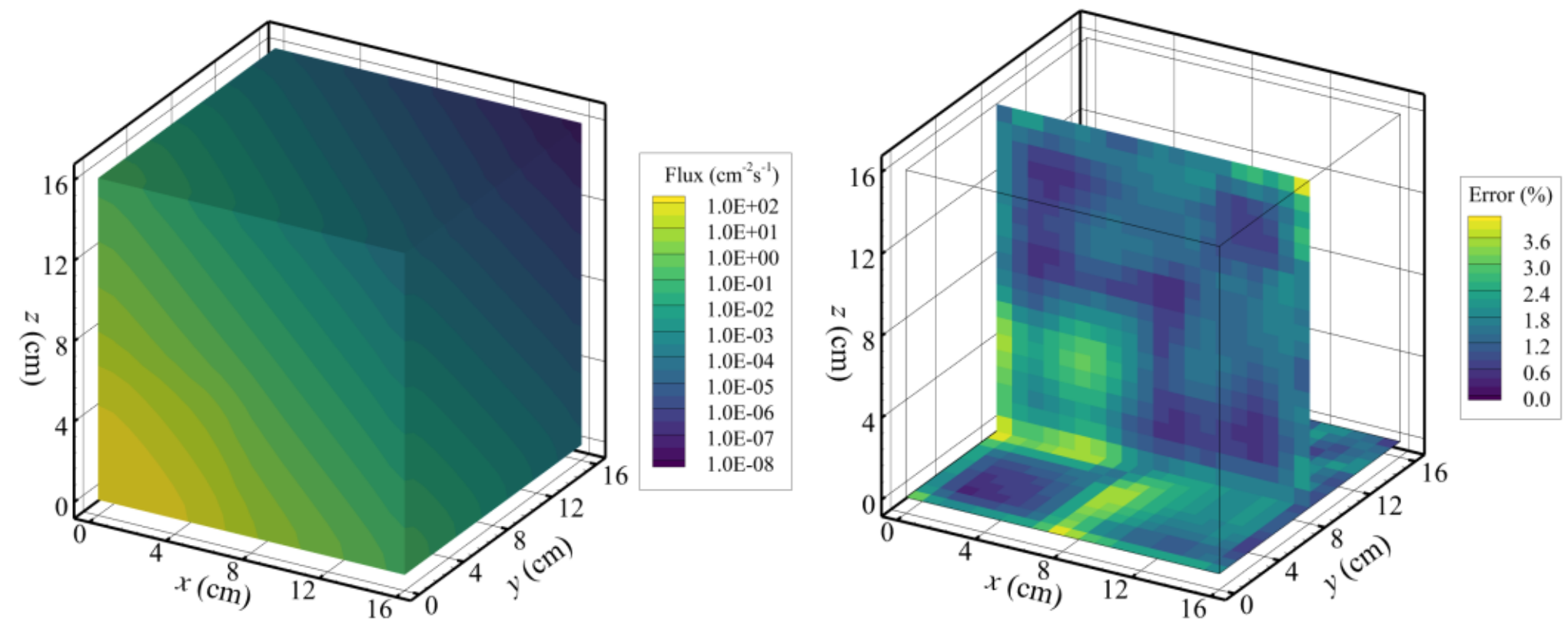

(a) $S_N$ using 0.8-cm grids

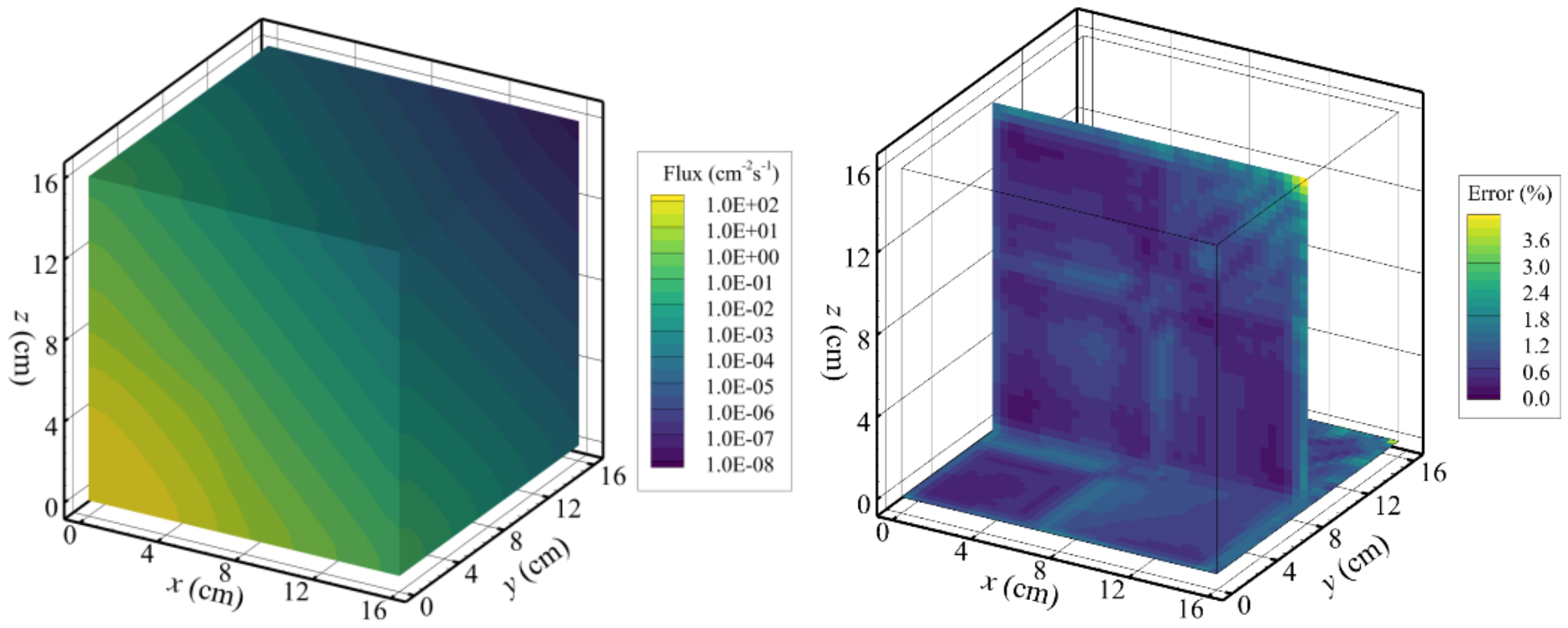

(b) $S_N$ using 0.4-cm grids

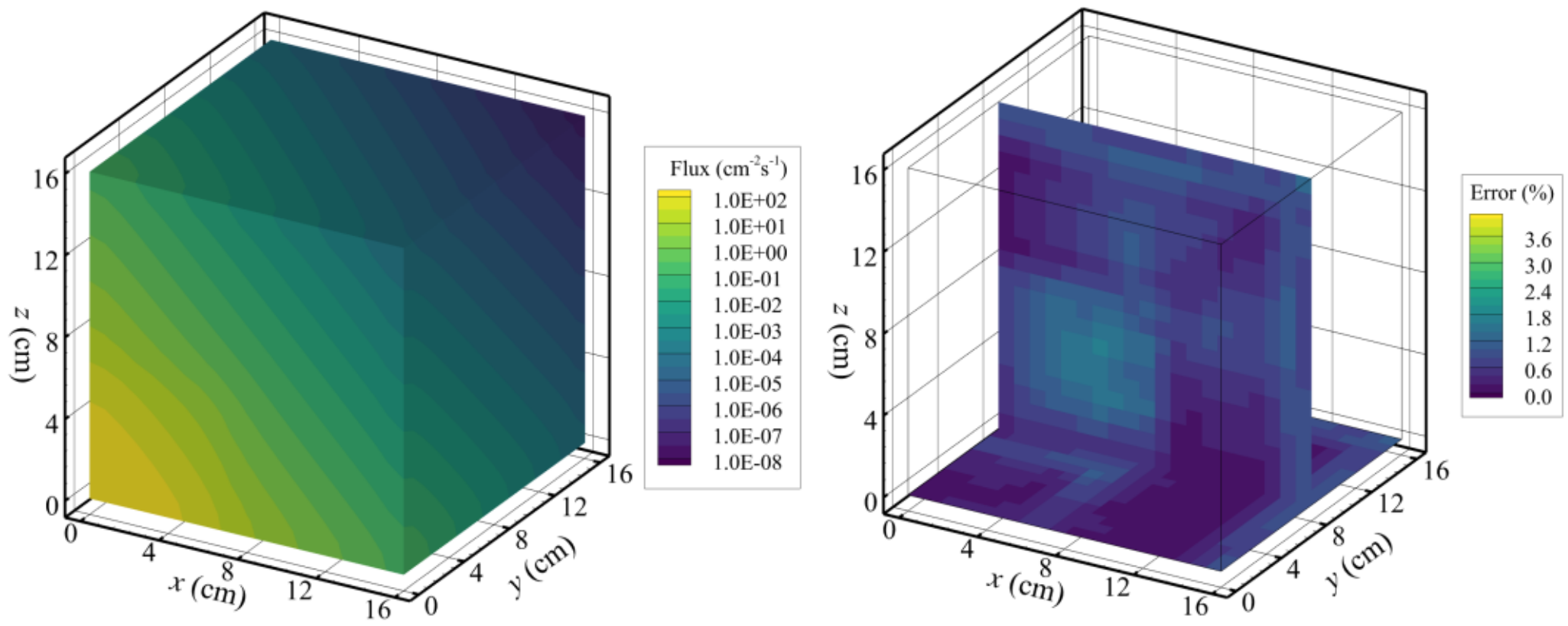

(c) pcNN-$S_N$ using 0.8-cm grids

Fig. 16. Results at the final time step for the 3-D time-dependent test problem

Fig. 17 shows $E_{RMS}$ norms at each time step for the $S_N$ calculation using 0.4-cm grids and pcNN-$S_N$ calculation using 0.8-cm grids. Over time, the neutron flux gradually evolves toward a steady-state distribution field. The computational times of $S_N$ and pcNN-$S_N$ simulations are 75 seconds and 24 seconds respectively, with our algorithm saving over 65% of the time.

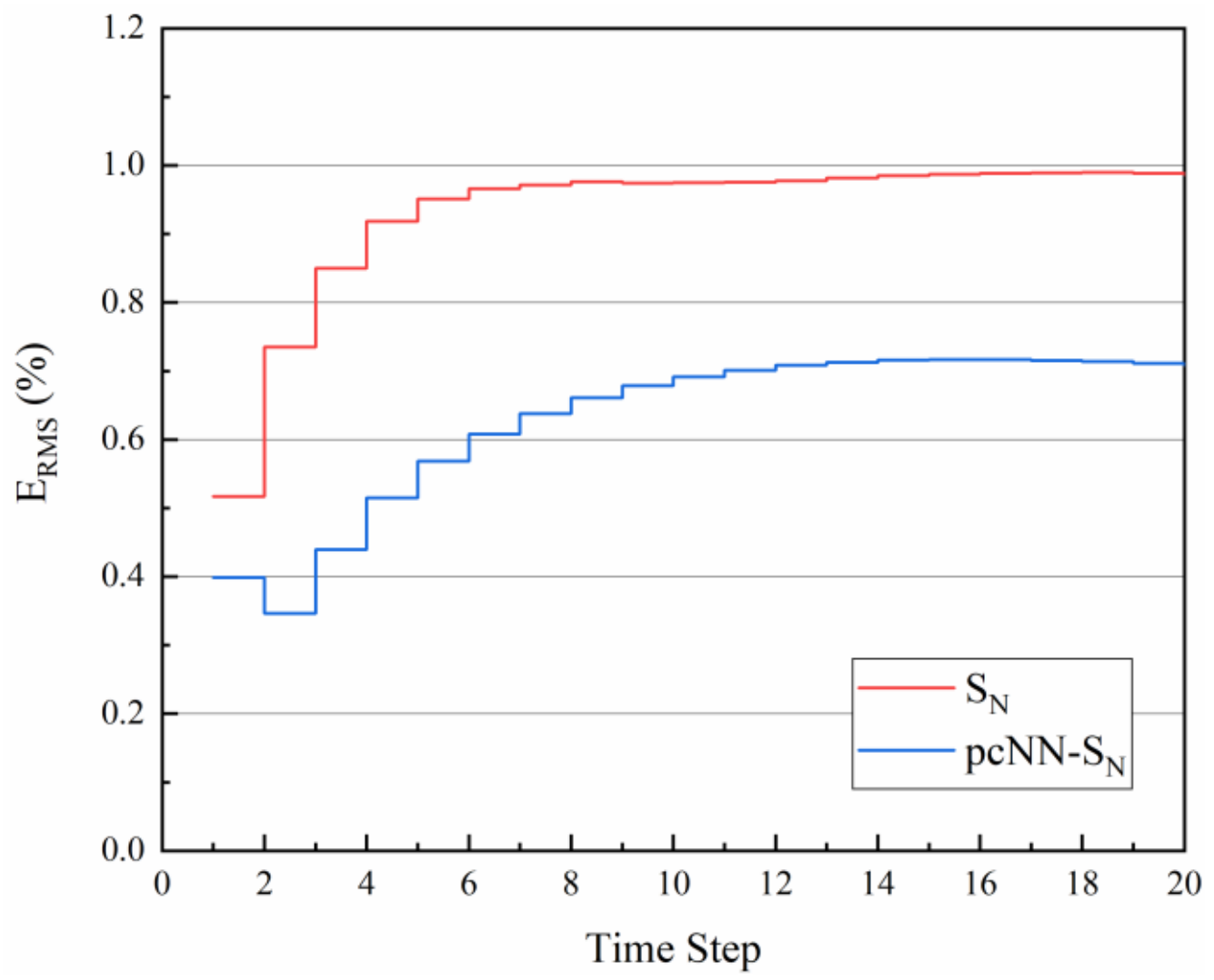

Fig. 17. $E_{RMS}$ norms at each time step for the 3-D time-dependent test problem

## 7. Conclusions

The high-dimensional nature of transport equations can result in high computational costs for traditional numerical simulations. Inspired by super-resolution machine learning methods, we explore a novel physics-enhanced coupling approach to improve $S_N$ transport calculations. One of the key aspects of super-resolution methods lies in identifying features in the low-accuracy solution that correspond to high-accuracy solution, and using these features to establish a mapping relationship between the two. We select the variables in the $S_N$ equation and the source reconstruction information at the subgrid scale as the feature information for the numerical solution.

Due to the high dimensionality of the input space and the wide distribution range of variables, network training is still quite challenging. We perform dimensionless transformation on the transport equations and use physically meaningful dimensionless quantities as local features. This treatment achieves dimensionality reduction of the network inputs and makes training easier. At the same time, the extrapolation of the original parameter set can be transformed into interpolation of dimensionless features, thereby enhancing the reusability of the network model. To satisfy physical conservation requirements, we incorporate the hard constraints of the balance equations into the network model and add soft constraints to the loss function. To address the permutation-equivariant property inherent in discretization scheme, we employ data augmentation approach to introduce additional data constraints during the training process. Based on numerical solutions from coarse and fine grids, we develop a predictor-corrector neural network to estimate correction factors for low-accuracy angular fluxes.

The network model is coupled with $S_N$ sweeps to perform online correction of numerical solutions, thereby forming a pcNN-$S_N$ coupling algorithm. We construct training problems to obtain simulation data and establish neural network models. By changing parameters such as geometric configurations, material cross sections, grid sizes, and boundary conditions, we generate new test problems to evaluate the performance of the coupling algorithm. Design of the physics-enhanced network inputs ensures that the network model possess strong reusability. Numerical results indicate that our algorithm can relax the requirements for grid spacing in the $S_N$ difference method, and the pcNN-$S_N$ results on coarse grids still agree well with high-accuracy solutions. The error norms of coupling algorithm are slightly lower than the traditional $S_N$ results obtained after halving the grid size. The network evaluation cost is lower than the cost introduced by mesh refinement, and the pcNN-$S_N$ algorithm achieves overall computational time savings exceeding 60%. Our machine learning method demonstrates potential for accelerating $S_N$ calculations.

This study explores a method coupling machine learning with traditional $S_N$ transport simulation, and we apply it to simple geometry problems. Further research is needed to develop a unified network model for strongly inhomogeneous media in complex problems. The output of our network can incorporate correction factors corresponding to subgrids to enhance computational resolution. In radiation shielding problems, the uncollided flux distribution is typically non-smooth and even discontinuous. Currently, we have not resolved the issue of network modeling for such data, which will be one of the key challenges for future research.

## CRediT Authorship Contribution Statement

Cong Liu: Methodology, Software, Formal analysis, Validation, Writing - Original Draft, Junxia Wei: Methodology, Resources, Writing - review & editing, Shuai Wang: Methodology, Writing - review & editing, Yong Heng: Methodology, Resources, Writing - review & editing.

## Declaration of Generative AI and AI-Assisted Technologies in the Writing Process

The authors used Grammarly and DeepL in order to improve the English language grammar and readability. The authors reviewed and edited the content as needed and take full responsibility for the content of the publication.

## Declaration of Competing Interests

The authors declare that they have no known competing financial interests or personal relationships that could have appeared to influence the work reported in this paper.

## Acknowledgements

The authors would like to thank all the members of AI++ team for their help and fruitful discussions. This work was supported by the National Natural Science Foundation of China (Grant Nos. 12331010, 12205020), and by the NSAF Joint Fund (Grant No. U2230208).

## Data Statement

The data that support the findings of this study are available from the corresponding author upon reasonable request.